\documentclass[12pt]{article}
\usepackage{graphics,cite,amssymb,epsfig,float,psfrag}
\usepackage[usenames,dvips]{color}
\usepackage{rotating}
\usepackage{axodraw}

\begin{document} 
 
\newcommand{\lsim}{\raisebox{-0.13cm}{~\shortstack{$<$ \\[-0.07cm] $\sim$}}~} 
\newcommand{\gsim}{\raisebox{-0.13cm}{~\shortstack{$>$ \\[-0.07cm] $\sim$}}~} 
\newcommand{\ra}{\rightarrow} 
\newcommand{\lra}{\longrightarrow} 
\newcommand{\ee}{e^+e^-} 
\newcommand{\gam}{\gamma \gamma} 
\newcommand{\nn}{\noindent} 
\newcommand{\non}{\nonumber} 
\newcommand{\beq}{\begin{eqnarray}} 
\newcommand{\eeq}{\end{eqnarray}} 
\newcommand{\s}{\smallskip} 
\newcommand{\mkk}{M_{\rm KK}} 
\newcommand{\vkk}{V_{\rm KK}} 
\def\NPB{Nucl. Phys. B} 
\def\PLB{Phys. Lett. B} 
\def\PRL{Phys. Rev. Lett.} 
\def\PRD{Phys. Rev. D} 
\def\ZPC{Z. Phys. C} 
\baselineskip=16.5pt 
%%%%%%%%%%%%%%%%%%%%%%%%%%%%%%%%%%%%%%%%%%%%%%%%%%%%%%%%%%%%%%%%%%%%%%%% 
 
\rightline{LPT--Orsay 07-45} 
\rightline{June 2007}

\vspace{0.7cm} 
 
\begin{center}

{\Large {\bf Kaluza--Klein excitations of gauge bosons at the LHC}} 
 
\vspace{0.7cm} 
 
{\large Abdelhak Djouadi$^1$, Gr\'egory Moreau$^1$, Ritesh K. Singh$^{1,2}$} 
 
\vspace{0.7cm}

$^1$ Laboratoire de Physique Th\'eorique, CNRS and Universit\'e Paris--Sud, \\ 
B\^at. 210, F--91405 Orsay Cedex, France. 
 
$^2$ Laboratoire de Physique Th\'eorique, LAPTH, F--74941 Annecy--le--Vieux, 
France.

\end{center} 
 
\vspace{0.5cm}  
\begin{abstract}   

We consider the Randall-Sundrum extra dimensional model with fields propagating
in the bulk based on an extended electroweak gauge symmetry with specific 
fermion charges and localizations that
allow to explain the LEP anomaly of the forward--backward
asymmetry for $b$--quarks, $A_{FB}^b$. We study the manifestations of
the strongly--interacting and electroweak gauge boson Kaluza--Klein excitations 
$\vkk$ at the LHC, with
masses of the order of a few TeV, which dominantly decays into top and bottom
quark pairs.  We first analyze the two--body tree--level production processes
$pp  \to t\bar t$ and $b\bar b$ in  which the Kaluza--Klein (KK) excitations 
of gauge bosons are exchanged. We
find that the additional channels can lead to a significant excess of events
with respect to the Standard Model prediction; characteristic top quark
polarization and angular asymmetries are quantitatively studied and turn out 
to probe the chiral structure of couplings to excited states. 
We then analyze higher order production processes for
the gauge boson excitations which have too weak or no couplings to light quarks
and, in particular, the loop induced process $gg \to \vkk \to t\bar t$  and
$b\bar b$ 
in which the anomalous $gg \vkk $ four--dimensional vertex has to be
regulated. 
The RS effects in this process, as well as in the four--body reactions $pp \to t\bar t
b\bar b$, $t\bar t t\bar t$, $b\bar b b\bar b$ and in the related 
three--body reactions $gb \to b t \bar t$, $b b\bar b$, 
in which the $\vkk$ excitations are mainly
radiated off the heavy quarks, are shown
to be potentially difficult to test at LHC, due to small phase space and low
parton density for $\mkk \gsim 3$ TeV.

\end{abstract}  
 
\newpage

\subsection*{1. Introduction}  
 
Among the extra--dimensional models which have been proposed in the recent years
as extensions of the Standard Model (SM), the one of Randall and Sundrum (RS)
\cite{RS}  seems to be particularly attractive. First, it addresses  the gauge
hierarchy problem without introducing a new energy scale in the fundamental
theory. Moreover, the variant in which  the SM fermions and gauge bosons are
propagating in the bulk allows for the unification of the gauge coupling
constants at high  energies \cite{UNI-RS} and provides a viable candidate of
Kaluza--Klein (KK) type for the Dark Matter of  the universe \cite{LZP2}.  This
version of the RS model with bulk matter offers also the possibility,  through a
simple geometrical mechanism, of generating the large mass hierarchies
prevailing  among SM fermions  \cite{RSloc,HSquark,GNeubertA}. Indeed, if the
various fermions  are placed along the extra dimension,  their different wave
functions overlap with the Higgs boson, which remains  confined on the
so--called TeV--brane for its mass to be protected, generate hierarchical
patterns among the effective four--dimensional Yukawa couplings.\s 

If this extra--dimensional model is to solve the gauge hierarchy problem, the
masses of the first KK excitations of the SM gauge bosons, $\mkk$, must be in
the vicinity of the TeV scale.
The direct experimental search for KK gluon excitations at Tevatron Run II 
leads to the bound $\mkk \gtrsim 800$ GeV \cite{Sridhar}. 
The high--precision measurements 
impose strong lower bounds on $\mkk$ as the exchanges of the KK excitations
lead to new and unacceptably large contributions to the electroweak observables
\cite{Burdman}. Nevertheless, it was shown \cite{ADMS} that if the SM gauge
symmetry is enhanced to the left--right  structure ${\rm SU(2)_L\! \times\!
SU(2)_R\! \times\! U(1)_{\rm B-L}}$, with $B$ and $L$ being respectively the
baryon and lepton numbers, the  high--precision data can be nicely fitted while
keeping  the masses down to an acceptable value\footnote{The severe indirect
bounds on $\mkk$  from precision data could also be softened down to a few TeV
in scenarios with brane--localized kinetic terms for fermions and gauge bosons
\cite{EWB-BraneFB}.} of $\mkk \simeq 3$ TeV, this lower limit being quite 
independent of the geometry of the extra dimension \cite{DelgFalk}.\s

Instead of the ${\rm U(1)_{\rm B-L}}$ group in the extended gauge symmetry as
originally proposed  in Ref.~\cite{ADMS}, other ${\rm U(1)_{X}}$ groups
with different  fermion charges can be considered \cite{AFB-Rold,RSAFB}.  An
important motivation is that  with specific charges of the new Abelian group and
for specific fermion localizations, different from the ones  considered  in
Ref.~\cite{ADMS}, the three standard discrepancy between the forward--backward
asymmetry $A_{FB}^b$ in $Z$ decays to bottom quarks  measured  at LEP and
elsewhere \cite{ref:LEP1,ref:PEPTRIS} and the SM prediction \cite{AFB-SM} is
naturally resolved,  while keeping all the other electroweak precision
observables, including the partial decay width of the $Z$ boson into $b$--quarks
$R_b \equiv \Gamma (Z \to b\bar{b}) / \Gamma (Z \to {\rm hadrons})$, unaffected
and in agreement with the data. More precisely, in the framework presented  in
Refs.~\cite{AFB-Rold,RSAFB},  the measured  values of the asymmetry
$A_{FB}^b(\sqrt{s})$ can be reproduced both at LEP1 \cite{ref:LEP1}, i.e. on
the  $Z$ pole, and at energies outside the $Z$--pole \cite{ref:PEPTRIS}, while
with the charge assignments of ${\rm U(1)_{\rm B-L}}$, the measured value of
the  asymmetry cannot be reproduced at energies lower than LEP1; see 
Ref.~\cite{RSAFB} for a detailed discussion.\s

The resolution of the $A_{FB}^b$ LEP anomaly is essentially due  to the fact that
since the third generation fermions should be localized  closer to the 
TeV--brane to get higher masses, their couplings to the KK gauge  bosons, which
are typically located near the TeV--brane, are larger and generate more
important   contributions to electroweak observables in the $b$ sector. One 
needs also to take into account the tension between generating a  large mass to
the top quark, which requires the left--handed doublet $(t,b)_L$  to be near the
TeV--brane, and satisfying the bounds from  precision data that constrain the
$Zb_L\bar b_L$  coupling to be almost SM--like, which forces the $(t,b)_L$  
doublet [via the mixing of the $Z$ boson with the KK states] to have small
couplings to the KK gauge bosons and thus to be far from the TeV--brane.
However, for a suitable choice  of the $(t,b)_L$ representation under ${\rm
SU(2)_R}$, the KK corrections to the  coupling $Zb_L\bar b_L$ can be suppressed
\cite{AFB-Rold,Csaki}. Alternatively,   the tension is softened in the framework
of Ref.~\cite{RSAFB} summarized  above, in which the $Zb_L\bar b_L$ and
$Zb_R\bar b_R$ couplings are chosen so that the  electroweak precision data are
reproduced\footnote{Note that there are also constraints on the top and bottom
quark couplings from Flavour Changing Neutral Current (FCNC) processes,
but those can be satisfied for $M_{KK}$ values around the TeV scale
Ref.~\cite{FCNC-review,AgasheTH,AgasheEXP}.}. \s

Thus and, generically speaking, in the context of the RS model with SM fields  
in the bulk,  the KK gauge  bosons dominantly couple to heavy SM fermions as
they are localized toward the TeV--brane (this typical feature can only be avoided
in some particular situations \cite{Ledroit}). In this case, the processes involving
the third generation $b$ and $t$ quarks are those which are expected to be 
significantly affected  by the presence of the new vector states. In
Ref.~\cite{RSAFB,Sher}, the impact of the KK excitations of the electroweak gauge
bosons has been discussed in the context of heavy flavor production at high
energy $\ee$ colliders (ILC).
Because of the high precision which can be achieved in the measurement of the
production rates and polarization/angular asymmetries in the $\ee \to t \bar t,
b \bar b$ processes at the ILC, KK vector excitations  with  large masses
can be probed, provided that they have non--vanishing couplings  also to the
initial light leptons. However, because of the limited energy of the ILC, $\sqrt
s=0.5$--1 TeV, and the expected large $\vkk$  masses, $\mkk\sim$
a few TeV,  the particles are accessible only indirectly through their virtual
exchange.\s

At the Large Hadron Collider (LHC), the energy is principle sufficient to
produce directly the KK excitations of the gluons and the electroweak gauge
bosons with masses in the multi--TeV range. The most direct manifestation of
these KK states\footnote{Another possibility to probe this variant of the RS
scenario at LHC is to look  for the production of KK gravitons \cite{GravRSbulk}, 
or, the new quarks that are present in the model \cite{Servant}. 
In the present analysis, we do not discuss these possibilities and we
assume that the new fermionic states are too heavy and do not end up as decay
products of the KK gauge bosons.} would be their production in the  Drell--Yan
processes,  $pp \to \vkk=g^{(1)}, \gamma^{(1)}, Z^{(1)}, Z^{'(1)}$. However, in
contrast to the (standard) production at the LHC of extra $Z'$ bosons from Grand
Unified  Theories (GUTs) \cite{Zprime-papers} and to some universal
extra--dimensional models \cite{RizzoUniv}\footnote{In Ref.~\cite{RizzoUniv},
this mechanism has been studied at the LHC in RS models with  bulk matter but
with the hypothesis of a universal fermion profile [made in order to totally 
suppress FCNC effects] which are not compatible with an interpretation of
fermion mass  hierarchies through the wave function overlap mechanism.}, there
are  many difficulties in the present scenario. The first one is that, contrary
to the new $Z'$s of GUTs, the KK excitations present  in this model have rather
tiny couplings to the light fermions and  large  couplings to the heavy
quarks. On the one hand, the production cross sections in the Drell--Yan process
$q \bar q \to \vkk$ are much smaller, the probability of finding a heavy quark
in the proton being tiny; in fact, in the case of the excitations of the extra
$Z'$, which is a superposition of the $Z_R$ and $Z_X$ bosons of the extended
${\rm SU(2)_R \times U(1)_X}$ group, the couplings to light fermions are almost
negligible and Drell--Yan production does not occur [except through $b\bar b \to
Z^{'(1)}$ which gives a small contribution due to the reduced $b$--quark density in the
proton]. On the  other hand, because of their large couplings to heavy
fermions, the new KK excitations will decay almost exclusively into $t\bar t$
and $b \bar b$ final states; these topologies have a less striking signature
than the GUTS $Z' \to \ell^+ \ell^¯$ with  $\ell =e, \mu$  channel and are
subject to a much more severe background at the LHC.  Another difficulty, that
is related to the previous one, lies in the fact that because of the  strong
couplings to heavy fermions, the KK excitations  have large total decay widths
and are not anymore narrow resonances. This will complicate the searches for
these states since in the processes $p \bar p \to \vkk \to t\bar t$ or $b\bar
b$,  the corresponding QCD backgrounds become also large as they have to be
integrated in wider bins~\cite{AgasheLHC,rmg}. \s 

There are two other possibilities of observing the heavy KK states. The first
one is single $\vkk$ production  in the gluon--gluon fusion mechanism $gg \to
\vkk$ which is  mediated by heavy $Q=t,b$ quark loops\footnote{The process $gg
\to V_{KK} \to t\bar t$ has been first suggested in Ref.~\cite{ADMS}
and recently mentioned in Ref.~\cite{Wang}; however,
because of  Yang's theorem the vector part of the $gg\vkk$ coupling will give
zero contribution while the axial part of the coupling will give a finite
contribution only if the $\vkk$ state is virtual and, thus, if its total width
is included.}; because of the chirality dependence of 
$\vkk tt$  and  $\vkk bb$ four--dimensional effective couplings, the
one--loop effective $gg \vkk $ vertex is anomalous and the chiral anomaly has to
be regulated by considering the St\"uckelberg mixing term
(for a review on the topic, see e.g. Ref.~\cite{Scrucca}). 
The second possibility is  associated production with
heavy quark pairs, $pp \to gg/q  \bar q \to Q\bar Q \vkk $ with  $Q=t,b$,
leading to $b \bar b t \bar t, b \bar b b \bar b$ and $ t\bar t t \bar t$ final
states; another mechanism, which is related to the later class of processes 
would be the two--body process $gb \to b\vkk  \to b b\bar b, b t\bar t$ where
the initial bottom parton is taken from the proton sea\footnote{ This second
class of processes have also been discussed in Ref.~\cite{Han-Valencia} for
rather general scalar and vector resonances which couple to top and bottom
quarks with the same strength.}.  In fact, when the couplings of the KK states
to light quarks are  small and the Drell--Yan process is not effective, these
mechanisms are the only  possibilities which give access to the new states; this
is the case for the production of the KK excitation of the new $Z'$ boson for
instance.  However, because  these processes are of higher order in perturbation
theory, the production cross sections are not substantial despite the fact that 
the couplings which are involved are large.  Nevertheless, in view of the 
interesting final state signature, a better control of the corresponding QCD
backgrounds might be possible.  \s 

In the present paper, we perform a complete study of the Drell--Yan and  gluon
fusion processes $q\bar q, gg \to b \bar b$, $t \bar t$ as well as the
associated production  $q\bar q/gg \to b \bar b t \bar t$, $b \bar b b \bar b, t
\bar t t \bar t$ and $gb \to bb\bar b, bt\bar t$ processes  at the LHC, in the
RS model with bulk matter. We will base our analysis on the framework which
resolves the $A_{FB}^b$ anomaly \cite{RSAFB}, but the results that we obtain can
be easily generalized to other scenarios.  In a parton--level analysis which
includes only the dominant QCD backgrounds, we show that for a set of
characteristic points of the parameter space,  the exchange of KK gauge bosons
[excitations of the gluon as well as of the electroweak bosons $\gamma$, $Z$ 
and new $Z'$] can lead to visible deviations with respect to the SM production
rates in the Drell--Yan   process $pp \to \vkk \to b \bar b, t\bar t$. For the
latter reaction, the signal rates are large and some top polarization and
asymmetries allow to test the chiral couplings to the KK modes.   In the case
where the KK excitations have too small couplings to the light quarks, the  loop
induced gluon fusion mechanism $gg \to t\bar t$ and $b\bar b$  as well  as the
associated production mechanisms $pp \to Q\bar Q \vkk  \to b \bar b t \bar t$
and $gb \to b t\bar t$, with an excess in the top or bottom  pair invariant mass
distributions can be in principle exploited. However, because of the large mass
of the KK excitations assumed in this study, the signal significance turns out
to be small, except potentially, for the $gb \to b t\bar t$ process. \s

Note that while the present analysis was on--going, the on--shell production of KK
gluons decaying into top quarks, $p p \to g^{(1)} \to t \bar t$ has been studied
at the LHC \cite{AgasheLHC,Wang} in the framework developed in 
Ref.~\cite{ADMS} with the
left--right extended gauge structure. The $ pp \to b \bar b$ process has not
been considered as it is expected to be less significant  than in the framework
\cite{RSAFB} considered here  where the right--handed $b$ quark is closer to the
TeV--brane,  with an increased coupling to KK gauge bosons; also in this case, 
the cross section for $t \bar t$ production  would be different as $b$ coupling
variations modify e.g. the KK gluon total width  as well as the $b \bar b$
parton initial state contribution (especially as here $b_R$
is closer to the TeV--brane so that it can get a non--vanishing coupling to the
$Z'$). Note also that the production of the
excitations of the electroweak gauge bosons as well as higher order processes
(including 3 and 4 quark final states)
have not been considered in Ref.~\cite{AgasheLHC,Wang}
(in \cite{Wang} the emphasis was put on the determination of excitation 
spin and methods of highly energetic top identification).\s

The paper is organized as follows. In the next section, we briefly describe the
theoretical framework in which we will work and the considered scenarios which
will be used for illustration; the decays of the KK states will also be
summarized. In section 3, the signal of the KK excitations at the LHC in the
Drell--Yan processes $pp \to t\bar t, b\bar b$ and the corresponding irreducible
QCD backgrounds are analyzed; relevant polarization and angular asymmetries are
studied in details. In section 4, we discuss the gluon fusion mechanism  $gg \to
\vkk  \to t\bar t, b\bar b$ and briefly the associated production processes with
$t\bar t/b\bar b$ pairs and those with a $b$-quark, as well as their
corresponding QCD backgrounds.  A brief conclusion will be given in section 5.

\subsection*{2. Physical framework}
\label{sub:context}

\subsubsection*{2.1 The specific RS extension}
\label{sub:theory}

In this paper, we consider the RS model in which SM fields propagate along the
extra spatial dimension, like gravity, but the Higgs boson remains confined on the
TeV--brane. 
In the RS scenario, the warped extra 
dimension is compactified over a $S^{1}/\mathbb{Z}_{2}$ 
orbifold. 
While the gravity scale on the Planck--brane is $M_{P}= 2.44\times
10^{18}$ GeV, the effective scale on the TeV--brane, $M_{\star}=e^{-\pi
kR_{c}} M_{P}$, is suppressed by a warp factor which depends on the curvature
radius of the anti--de Sitter space $1/k$ and the compactification radius $R_c$.
The product $k R_{c} \simeq 11$ leads to $M_{\star}\!=\!{\cal O}(1)$ TeV, thus
addressing the gauge hierarchy problem.\s

The fermion mass values are dictated by their wave function localization.  In
order to control these localizations, the five--dimensional fermion fields
$\Psi_{i}$, with $i=1,2,3$ being the generation index, are usually coupled to
distinct masses $m_{i}$ in the fundamental theory. If $m_{i}= {\rm sign}(y)
c_{i} k$, where $y$ parameterizes the fifth dimension and $c_{i}$ are
dimensionless parameters, the fields decompose as $\Psi _{i}(x^{\mu },y)=
\sum_{n=0}^{\infty }\psi_{i}^{(n)}(x^{\mu }) f_{n}^{i}(y)$, where $n$ labels
the tower of KK excitations and $f_{0}^{i}(y)=e^{(2-c_{i})k|y|} / N_{0}^{i}$
with $N_{0}^{i}$ being a normalization factor. Hence, as $c_i$ increases, the
wave function $f_{0}^{i}(y)$ tends to approach the Planck--brane at $y=0$.\s

In order to protect the electroweak observables against large deviations and, 
at the same time, to resolve the anomaly in the forward--backward asymmetry for
$b$--quark production $A_{FB}^b$, the electroweak gauge symmetry is enhanced to
${\rm SU(2)_L \times SU(2)_R \times U(1)_{X}}$ with the following fermion
representations/charges under the group gauge \cite{RSAFB,AFB-Rold}:
\begin{eqnarray}
Q_L^i &\equiv& ({\bf 2},{\bf 1})_{\frac{1}{6}}; \ \ \
u_R^i \in ({\bf 1},{\bf 2})_{\frac{1}{6}} \ \mbox{with} \ I_{3R}^{u^i_R}=+
\frac{1}{2}; \nonumber \\
&&d_R^i \in ({\bf 1},{\bf 2})_{-\frac{5}{6}} \ \mbox{with} \ I_{3R}^{d^i_R}=+
\frac{1}{2}
\nonumber \\
L_L^i &\equiv& ({\bf 2},{\bf 1})_{-\frac{1}{2}}; \ \ \
\ell_R^i \in ({\bf 1},{\bf 2})_{-\frac{1}{2}} \ \mbox{with} \
I_{3R}^{\ell^i_R}=-\frac{1}{2}
\label{eq:scen}
\end{eqnarray}
respectively for the left--handed ${\rm SU(2)_L}$ doublets of quarks,
right--handed up/down type quarks, doublets of leptons and right--handed
charged leptons. The usual symmetry of the SM is recovered after the breaking
of both ${\rm SU(2)_{R}}$ and ${\rm U(1)_{X}}$ on the Planck--brane, with
possibly a small breaking of the ${\rm SU(2)_R}$ group in the bulk. Note the
appearance of a new $Z'$ boson [but without a zero--mode] which is a
superposition of the state $\widetilde W^3$ associated to the ${\rm SU(2)_R}$
group and $\widetilde B$ associated to the ${\rm U(1)_{X}}$ factor; the
orthogonal state is the SM hypercharge $B$ boson. Up--type quarks acquire
masses through a Yukawa interaction of type $\overline{({\bf 2},{\bf 1})}_{
\frac{1}{6}} ({\bf 2},{\bf 2})_{0}({\bf 1},{\bf 2})_{\frac{1}{6}}$, the Higgs
boson being embedded in a bidoublet of ${\rm SU(2)_L \times SU(2)_R}$. Because
of their specific ${\rm SU(2)_R}$ isospin assignment $I_{3R}^{d^i_R}$,
down--type quarks become massive via another kind of invariant operator 
as in Ref.~\cite{RSAFB,AFB-Rold}.

\subsubsection*{2.2 EW constraints and the fermion couplings}
\label{sub:parspace}

From the theoretical point of view, solving the gauge hierarchy problem forces
the masses of the first KK excitations of the SM gauge bosons, $\mkk=M_{
\gamma^{(1)}}=M_{g^{(1)}} \simeq M_{Z^{(1)}} \simeq M_{Z^{\prime (1)}}$, to be
of order the TeV scale. Indeed, one has $\mkk=2.45 k M_{\star}/M_{P} \lesssim
M_{\star} = {\cal O}$(TeV) since the theoretical consistency bound on the
five--dimensional curvature scalar leads to $k<0.105 M_{P}$.  More precisely,
the maximal value of $\mkk$ is fixed by this theoretical consistency bound and
the $kR_{c}$ value. One could consider a maximal value of $\mkk \simeq 10$ TeV
which corresponds to $kR_{c}=10.11$.  From the experimental point of view, as
mentioned previously, the electroweak (EW) high--precision data force the KK mass  to
be larger than $\mkk \sim 3$ TeV.  Given these theoretical and experimental
considerations, we will fix the masses of the KK excitations to a common value  
$\mkk= 3$ TeV in the present study.\s

With light SM fermions [leptons and first/second generation quarks]
characterized by $c_{\rm light}>0.5$, $c$ being the parameter which determines
the fermion localization as discussed in the subsection above, the bulk
custodial isospin gauge symmetry insures an acceptable fit of the oblique
corrections to electroweak observables for $\mkk \gtrsim 3$ TeV
\cite{ADMS}. The large value of the top quark mass requires $c_{t_R}<0.5$ and
$c_{Q^3_L}<0.  5$, with $c_{Q^3_L}\!=\!c_{t_L}\!=\!c_{b_L}$ [as the states $b_L$
and $t_L$ belong to the same ${\rm SU(2)_L}$ multiplet] so that the top and
bottom quarks have to be treated separately in the studies of the electroweak
precision data. In the framework developed in Ref.~\cite{RSAFB}, the precision
data in the $b$ sector, that is, the $b$--quark forward--backward asymmetry
$A_{FB}^b$ and the $b\bar b$ partial decay width of the $Z$ boson $R_b \equiv
\Gamma (Z \to b\bar{b}) / \Gamma (Z \to {\rm hadrons})$, are correctly
reproduced with $\mkk=3$ TeV for certain values of the $c$ parameters
depending on the coupling constant $g_{Z'}$ of the new $Z'$ boson.\s

For instance, if the coupling $g_{Z'}$ is equal to $0.6 \sqrt{4 \pi}$, the best
fit of the observables $R_b$ measured at LEP1 and $A_{FB}^b(\sqrt{s})$ measured
at various center--of--mass energies is obtained for $c_{Q^3_L} \simeq 0.36$ and
$c_{b_R} \simeq 0.438$ [the fit corresponds to $\chi^2_{\rm RS}  \simeq 14$ and
substantially improves the one in the SM, $\chi^2_{\rm SM} \simeq 24$]. For this
$c_{b_R}$ value, one obtains the values $Q(c_{b_R}) \simeq 0.55$ and
$Q'(c_{b_R})\simeq 0.75$, where $Q(c)$ ($Q'(c)$) is the ratio of the
four--dimensional effective coupling between the $g^{(1)}/\gamma^{(1)}/Z^{(1)}$
($Z^{\prime (1)}$) boson and the SM fermions, over the coupling of the 
gluon/photon/$Z$ (would be $Z'$) boson zero--mode. In fact, the value chosen
above for the $g_{Z'}$ coupling is close to the typical limit obtained from the
perturbativity condition, $2 \pi k R_c g_{Z'}^2 / 16 \pi^2 < 1$, for the
coupling of the KK excitation of the $Z'$ boson, when quantum corrections are
included. Indeed, the effective four-- dimensional coupling of the $Z^{\prime
(1)}$ boson to SM fields is increased by a factor as large as $\sqrt{2 \pi k
R_c}$ relatively to $g_{Z'}$ for SM fields near the TeV--brane, like for example
the Higgs boson\footnote{A bound of the same order can be derived from
considerations on the strong coupling regime of the five--dimensional theory:
perturbativity imposes that the five--dimensional loop expansion parameter,
estimated at an energy $k$ to be $\sim (g_{Z'}^{5D})^2 k / 16 \pi^2 = \pi k R_c
g_{Z'}^2 / 16 \pi^2$, should be smaller than unity \cite{MCHM}.}.\s

For the choice $g_{Z'}=0.3 \sqrt{4 \pi}$, the best fit of $R_b$ and
$A_{FB}^b(\sqrt{s})$ [still corresponding to a good fit of the data,
$\chi^2 \simeq 14$] is obtained for the $c$ values $c_{Q^3_L} \simeq 0.36$ and
$c_{b_R} \simeq 0.135$, leading to the charges $Q(c_{b_R}) \simeq 3.0$ and
$Q'(c_{b_R}) \simeq 3.2$. For this value of $c_{Q^3_L}$, one obtains for
instance a top quark mass $m_t \simeq 90$ GeV or $m_t \simeq 80$ GeV with,
respectively, 
$c_{t_R} \simeq -0.3$ or $c_{t_R}\simeq +0.1$, which are in the correct order of
magnitude. In order to determine the set of parameters that would reproduce
exactly the measured $m_t$ value, one need to study the full three--flavour mass
matrix [by fixing also each Yukawa coupling constant] and include the effect of
mixing between $t$ and its KK excitations, which is beyond the scope of this
work.
For $c_{Q^3_L} \simeq 0.36$ and $c_{b_R} \simeq 0.438$, the bottom mass 
has also a correct order of magnitude. The bottom and top quark Yukawa
couplings have different gauge structures in the present context and 
the bottom Yukawa coupling constant could be taken smaller (while
choosing a smaller $c_{b_R}$, in order to still generate a satisfactory
bottom mass) as proposed in Ref.~\cite{AFB-Rold} in order to have 
a weak breaking of the custodial symmetry subgroup protecting the
$Z b_L \bar b_L$ coupling.\s

In this study, we will consider four typical scenarios which are defined by the
$c$ assignments for the top and bottom quarks and the value of the coupling
$g_{Z'}$; these parameters and their corresponding $Q(c)$ and $Q'(c)$ charges are
summarized in Table 1.  In the four points of parameter space, labeled
$E_{1,2,3,4}$, the assignment for the left--handed $(t,b)$ doublet, 
$c_{Q^3_L}=0.36$ which leads to the charges $Q(c_{Q^3_L})=1.34$ and
$Q'(c_{Q^3_L})=1.52$, has been adopted in order to reproduce the experimental
values of $A_{FB}^b$ and $R_b$. For the light fermions, i.e. the leptons and
the first and second generation quarks, we will set $c_{\rm light} \gtrsim 0.5$, 
as necessary to generate sufficiently small masses. This
leads to a  small $Q$ charge for these fermions, $Q(c_{\rm light}) \simeq -0.2$,
and a zero $Z'$ charge, $Q'(c_{\rm light})=0$. This means that the couplings of
the KK excitations of the gluon, photon and $Z$ boson have small  couplings to
light fermions compared to the top and bottom quarks [an order of magnitude
smaller in general], while the KK excitations of the $Z'$ boson do not couple to
these  light fermions at all.\s  

\begin{table}[!h]
\renewcommand{\arraystretch}{1.5}
\begin{center}
\begin{tabular}{|c|c|ccc|ccc|}\hline
    & $g_{Z'}$ & $c_{b_R}$ & $Q(c_{b_R})$ & $Q'(c_{b_R})$ 
               & $c_{t_R}$ & $Q(c_{t_R})$ & $Q'(c_{t_R})$\\ \hline 
$E_1$ & $0.6 \sqrt{4 \pi}$ & 0.438 & 0.55 & 0.75 & --0.3 & 4.90 & 5.01 \\
$E_2$ & $0.3 \sqrt{4 \pi}$ & 0.135 & 3.04 & 3.19 & --0.3 & 4.90 & 5.01 \\
$E_3$ & $0.6 \sqrt{4 \pi}$ & 0.438 & 0.55 & 0.75 & +0.1 & 3.25 & 3.40 \\
$E_4$ & $0.3 \sqrt{4 \pi}$ & 0.135 & 3.04 & 3.19 & +0.1 & 3.25 & 3.40 \\ \hline
\end{tabular}
\end{center}
\vspace*{-3mm}
\caption{The four scenarios labeled $E_{1,\dots,4}$ to be considered in this 
study and their respective $c$ assignments, $Q(c),Q'(c)$ charges and $g_{Z'}$
couplings. The other (common) parameters are $c_{Q^3_L}=0.36$, which leads
to $Q(c_{Q^3_L})=1.34$ 	and $Q'(c_{Q^3_L})=1.52$), and $c_{\rm light} \gtrsim 
0.5$, for which $Q(c_{\rm light})=-0.2$ and $Q'(c_{\rm light})=0$; $\mkk=3$ 
TeV.}
\label{tab:par}
\end{table}

\subsubsection*{2.3 Decays of the KK gauge bosons}
\label{sub:width}

The KK excitations of the gauge bosons will decay into pairs of SM fermions,
$\vkk  \to f \bar f$; as mentioned previously, we will assume that all the KK
excitations of the fermions, as well as the zero--modes of the new fermions 
that are present in the model, are too heavy, $2 m_{f}\gsim M_{\vkk }$, so that
they do not appear in the decays of the KK gauge bosons. The partial decay 
width of $\vkk $ into a fermion species $f$ is given by
\begin{eqnarray}
\Gamma (\vkk  \to f\bar f) &=& \frac{M_{\vkk } \beta_fC_f^{\vkk }}{24\pi}\left[
\frac{3+\beta_f^2}{4} \left( (g_{f_L}^{\vkk })^2 + (g_{f_R}^{\vkk })^2\right) 
+\frac{3 (\beta^2_f -1)}{2} \ g_{f_L}^{\vkk } g_{f_R}^{\vkk } \right]
\label{eq:width}
\end{eqnarray}
where $m_f$ is the mass of the fermion $f$, $\beta_f=\sqrt{1-4m_f^2/M_{\vkk }^2}$
its velocity in the rest frame of the KK gauge boson and $C_f^{\vkk }$ the  color
factor: $C_q^{\vkk }=3$, $C_\ell^{\vkk }=1$ for electroweak boson couplings to,
respectively,  quarks and leptons and $C_q^{g^{(1)}}=\frac12$, $C_\ell^{g^{(1)}}=0$ for
the coupling of the KK gluon to quarks and colorless fermions.  The left-- and
right--handed couplings of the KK gauge boson $\vkk $ to the fermion $f$ are
denoted by $g_{f_L}^{\vkk }$ and $g_{f_R}^{\vkk }$; in terms of the  couplings of
the corresponding zero--mode, $g_{f_{L/R}}^{V}$, they are given by
$g_{f_{L/R}}^{\vkk } = g_{f_{L/R}}^{V} Q^{(\prime)} (c_{f_{L/R}})$, with
$Q(c_{f_{L/R}})$ and $Q'(c_{f_{L/R}})$ as defined in the previous subsection and
given in Table \ref{tab:par}.  The total decay width of the KK excitation is simply
$\Gamma_{\vkk }= \sum_f \Gamma (\vkk  \to f\bar f)$, where the sum runs over all SM
fermions.\s  

Because the KK gauge boson couplings to fermions are directly proportional to
the charges $Q(c)$ and $Q'(c)$, the partial decay widths $\Gamma (\vkk  \to f\bar f)
\propto (g_{f_L}^{\vkk })^2 + (g_{f_R}^{\vkk })^2$ in the limit $M_{\vkk } \gg m_f$, 
are typically two
orders of magnitude larger in the case of final state top and bottom quarks than
for the light fermions.  The branching ratios  BR$(\vkk  \to t\bar t)$ and BR$(\vkk 
\to  b\bar b)$, with $\vkk =\gamma^{(1)}$, $Z^{(1)}$, $Z'^{(1)}$ and $g^{(1)}$, 
are displayed in Table 2 in the four scenarios $E_{1,..,4}$ introduced 
previously. As can be seen, the sum of these two branching ratios is close to
unity which means that the KK excitations decay almost exclusively into the
heavy $t,b$ quarks and that  little room is left for decays into light
quarks and  leptons [in fact, there is no room at all for these decays in the
case of the $Z'^{(1)}$ boson as the charge $Q'(c_{\rm light})$ is zero]. In our
analysis, we will therefore concentrate on the $t,b$  final state decay products
of the  KK gauge bosons.\s 

Finally, the total decay widths of the KK excitations $\vkk $ are also given in
Table 2 for the four scenarios. As they grow proportionally to the mass $M_{\vkk
}$, they are rather large for the value $\mkk=3$ TeV.  For instance,
the decay width of the KK gluon is of the order of a few hundred GeV and is
between 10\% and 20\%  of the ${g^{(1)}}$ mass; the KK state can be thus
considered as a relatively narrow resonance. Due to the significant
$g_{Z'}$ values taken, the total decay widths are 
smaller for the KK excitations of the photon and $Z$ boson than for the KK
excitation of the $Z'$ boson. In  the latter case, 
the total width is comparable to the $Z^{\prime (1)}$ 
mass and one can hardly consider the state as a true resonance.\s  

\begin{table}[!h]
\renewcommand{\arraystretch}{1.5}
\begin{center}
\begin{tabular}{|c|c|c|c|c|}\hline
    & $g^{(1)}$ & $\gamma^{(1)}$ & $Z^{(1)}$ & $Z^{\prime (1)}$ \\ \hline 
$E_1$  & 627/0.08/0.91  & 137/0.02/0.96   & 75/0.28/0.68   & 3360/0.02/0.98 \\ 
$E_2$  & 828/0.30/0.69  & 149/0.10/0.89   & 79/0.31/0.65   & 1080/0.29/0.71 \\
$E_3$  & 328/0.14/0.83  & 68.0/0.04/0.93  & 54/0.39/0.56   & 1580/0.04/0.96 \\
$E_4$  & 530/0.47/0.52  & 80.0/0.18/0.79  & 58/0.43/0.53   & 661.0/0.47/0.53 \\
\hline
\end{tabular}
\end{center}
%\vspace*{-3mm}
\caption{The total decay widths (in GeV) and the $b\bar b$ and $t\bar t$ 
branching ratios of the first KK excitations of the gluon, photon, $Z$ and 
$Z'$ bosons in the four selected scenarios $E_{1},.., E_4$ for $\mkk=3$ TeV: 
$\Gamma_{\vkk }$/BR($\vkk  \to b\bar b)$/BR($\vkk  \to t\bar t)$. }
\label{tab:br}
%\vspace*{-3mm}
\end{table}

\subsection*{3. Top and bottom quark pair production}

\subsubsection*{3.1 General features}

The most straightforward way to produce the KK excitations of the gauge 
bosons\footnote{From now on, we will restrict ourselves to the first KK 
excitations, $\vkk =\gamma^{(1)}$, $Z^{(1)}$, $Z'^{(1)}$ and $g^{(1)}$.} $\vkk $ 
at the LHC is via the Drell--Yan process, 
\beq
pp \to q\bar q \to \vkk  \to Q\bar Q \ , \ \ Q=t,b
\eeq
with the gauge bosons $\vkk $ subsequently decaying into top and bottom quarks; 
Fig.~1a. As discussed previously, $\vkk $ decays into the light fermion  states
[leptons and first/second generation quarks] will be neglected as their
branching ratios are very small in the framework that we are considering here.
The relatively  small couplings of the initial quarks $q \equiv u,d,s,c$ to 
$\vkk $ lead to  smaller production rates compared to, for instance, the
production of $Z'$  bosons from GUTs which have full--strength couplings to
light fermions. Because the $\vkk $ couplings to bottom quarks are large, the
partonic cross section from the subprocess $b\bar b \to  \vkk $ would, in 
principle, be more substantial than for the $q \bar q \to \vkk $ subprocesses
but, after folding with the smaller $b$--parton density, the corresponding
hadronic cross section is reduced. Note that since the KK gauge bosons have
different  couplings to left-- and right--handed fermions, one expects the
produced $t/b$   quarks to be polarized and to have a forward--backward
asymmetry.

\begin{figure}[!h]
\begin{center}
\vspace*{-.2cm}
\hspace*{-9.5cm}
\begin{picture}(300,100)(0,0)
\SetWidth{1.}
\SetScale{1.1}
\ArrowLine(100,75)(150,50)
\ArrowLine(100,25)(150,50)
\Photon(150,50)(200,50){4}{5}
\ArrowLine(200,50)(250,25)
\ArrowLine(200,50)(250,75)
\Text(110,70)[]{$q$}
\Text(110,40)[]{$\bar q$}
\Text(190,69)[]{$\vkk $}
\Text(275,70)[]{$Q$}
\Text(275,40)[]{$\bar Q$}
\Text(80,85)[]{\bf a)}
\hspace*{5cm}
\Text(170,85)[]{\bf b)}
\ArrowLine(180,75)(230,75)
\ArrowLine(230,75)(280,75)
\Photon(230,75)(230,25){4}{5}
\ArrowLine(180,25)(230,25)
\ArrowLine(230,25)(280,25)
\Text(199,70)[]{$b$}
\Text(199,40)[]{$\bar b$}
\Text(235,50)[]{$\vkk $}
\Text(300,70)[]{$b$}
\Text(300,40)[]{$\bar b$}
\end{picture}
\end{center}
\vspace*{-1.2cm}
\caption[]{\it Feynman diagram for heavy $Q=t,b$ quark pair production at 
the LHC through $q\bar q$ annihilation, where $q$ stands for initial state 
quarks and KK for the KK excitations of neutral gauge bosons, namely
$\gamma^{(n)}$, $g^{(n)}$, $Z^{(n)}$ and $Z^{\prime (n)}$  with $n=0,1,2,\dots$
being the KK level; mainly the zero--modes $n=0$, except for $Z'$, and the 
first KK excitations $n=1$ contribute to the signal. For bottom quark
production there is an additional $t$--channel diagram exchanging neutral KK 
gauge bosons with bottom partons in the initial state (b).
\protect\label{fig:ff}} 
\vspace*{-2mm}
\end{figure}
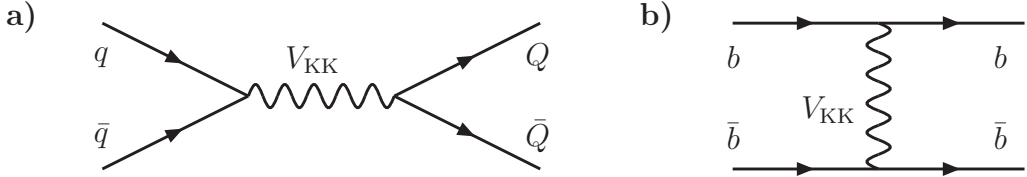

Because the KK excitations have substantial total decay widths [see Table 2], the
narrow width approximation in which the production and  decay processes are
factorized and considered separately is not sufficient.  Indeed, one needs to
consider the virtual exchange of the KK excitations in which the total width is
included in a Breit--Wigner form, together with the exchange  of the zero modes.
The full interference between the zero and first modes of  the gluon, on the one
hand, and of the electroweak bosons, on the other hand,  should be taken into
account [because of color conservation, there is no  interference between the
amplitudes in which the gluon and the electroweak  gauge bosons are exchanged].
For the $b \bar b$ production, the subprocesses are also initiated by bottom 
partons and one should also consider the channel in which the KK gauge bosons 
are exchanged in the $t$--channel; Fig.~1b.\s

The signal for the KK gauge bosons $q\bar q \to \vkk  \to  Q\bar
Q$ and  the main SM background $q \bar q \to  Q\bar Q$ 
%[for the background, one
%can safely ignore the exchange of the photon and $Z$ boson which leads to much 
%smaller contributions than in the gluon case, as we have numerically checked] 
and $gg  \to  Q\bar Q$ %[which 
%gives a much more substantial contribution as the gluon luminosity is more
%important at higher energies] 
have to be considered simultaneously.   For the
signal reaction, we have calculated the matrix element squared of the  process
$pp \to Q\bar Q$ with {\it polarized} final state quarks and incorporated  the
exchange, including the $t$--channel contributions, of all the SM gauge bosons
as well as their KK excitations and  those of the  $Z'$ boson. For the
background processes, the contributions of both the $q\bar q$ and $gg$ initial
states have been calculated  and the exchange of the KK excitations of the gauge
bosons were switched off. To obtain the cross  sections $\sigma$ at the hadronic
level, we use the CTEQ5M1 set of parton  distributions \cite{CTEQ} with the
factorization and a renormalization  scales set to the invariant mass of the
$Q\bar Q$ system, $\mu_F=\mu_R= m_{Q\bar Q}/2$.\s  

The significance ${\cal S}$ of the signal in the RS model can be then defined 
as
\begin{equation}
{\cal S}_{\cal L}  = \frac{ \sigma^{\rm RS+SM} - \sigma^{\rm SM}}{ \sqrt{ 
\sigma^{\rm SM}  } } \times \sqrt{ {\cal L}},
\label{eq:signif}
\end{equation}
where ${\cal L}$ denotes the total  integrated LHC  luminosity. In our analysis,
we will chose two examples for the luminosity: a lower value ${\cal L}=10$ 
fb$^{-1}$ that is expected in the first years of the LHC running and  a high
value ${\cal L} =100$ fb$^{-1}$ which is expected to be reached after a  few 
years of running.\s 

In order to enhance the signal, which is peaked at the invariant mass of the $Q\bar Q$ 
system (as will shown in details later), 
and to suppress the continuum background, one needs to select events 
near the KK resonance. Throughout this analysis, we impose a cut on the 
invariant mass  of the $Q \bar Q$ final state
\beq
|m_{Q\bar Q}-\mkk| \le \Gamma_{\vkk} 
\label{eq:qqcut}
\eeq
To further reduce the backgrounds, we  also impose the following cuts on
the transverse momenta of the two final jets and their rapidity 
\beq 
p_T^{Q,\bar Q} \ \ge \ 200\ {\rm GeV} \ , \  |\eta_{Q,\bar Q}| \ \le \ 2
\label{eq:ptcut}
\eeq
as in the signal, the $p_T$ of the jets is peaked close to $\frac12 M_{\vkk }$ and
the production is central, while in the background, the jets are peaked in the
forward and backward directions and the bulk of the cross section is for low
$p_T$ jets. These cuts can certainly be improved to optimize the signal to
background ratio but we will refrain from performing such an optimization here.
In  fact, in this preliminary and simple parton--level investigation, we will
simply compare the signal and the main corresponding physical background to
determine if, grossly,  the exchanges of the KK states could give rise to a
promising and possibly  detectable signal. A more   detailed Monte--Carlo study,
including precise experimental effects,
initial and final state radiation, the correct identification of the
$t/b$ states and the measurement of their  momenta,  more efficient and 
realistic cuts and detection efficiencies, which is beyond our scope 
here, will be needed.
Nevertheless, the large significances that we will present are expected to remain at 
an interesting level after implementation of these experimental effects.\s

\subsubsection*{3.2 The cross sections}

\begin{figure}[!h]
\begin{center}
\epsfig{file=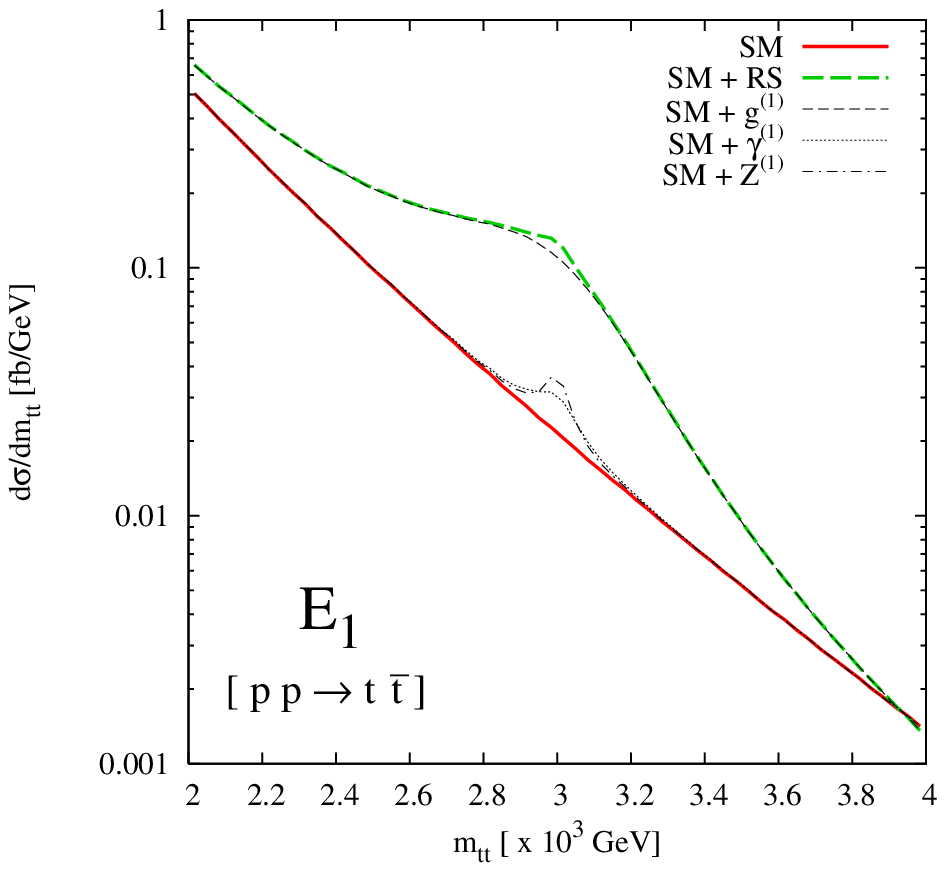,width=8.0cm}
\epsfig{file=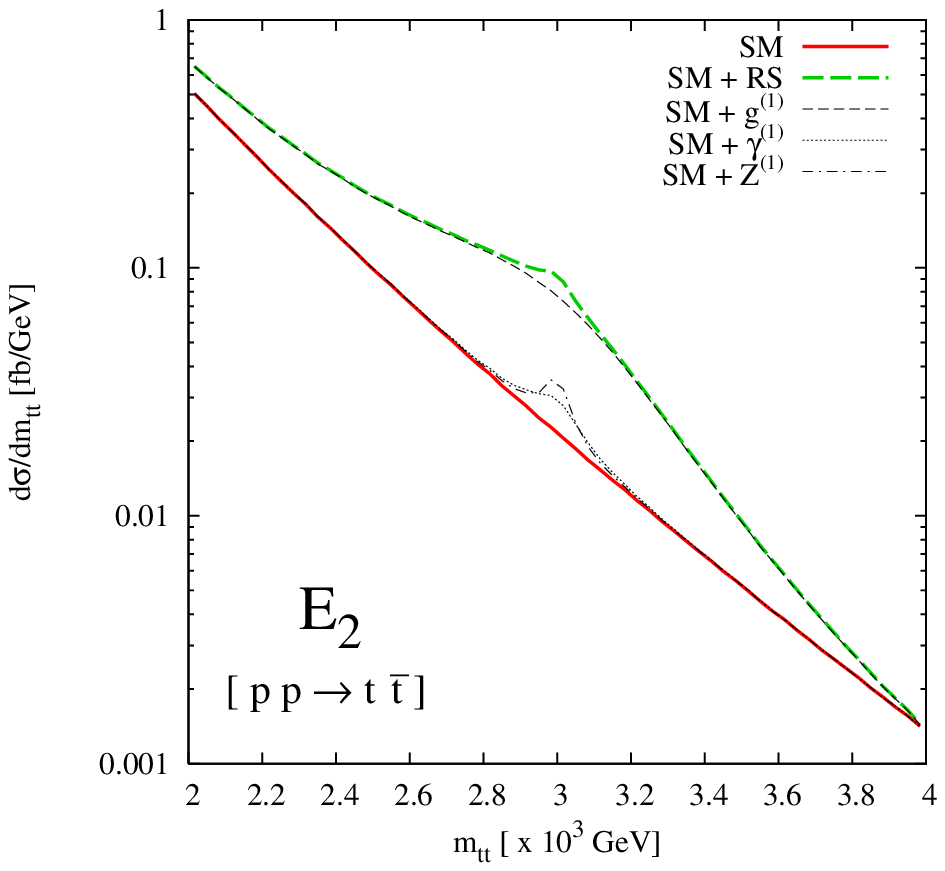,width=8.0cm}\vspace*{5mm}
\epsfig{file=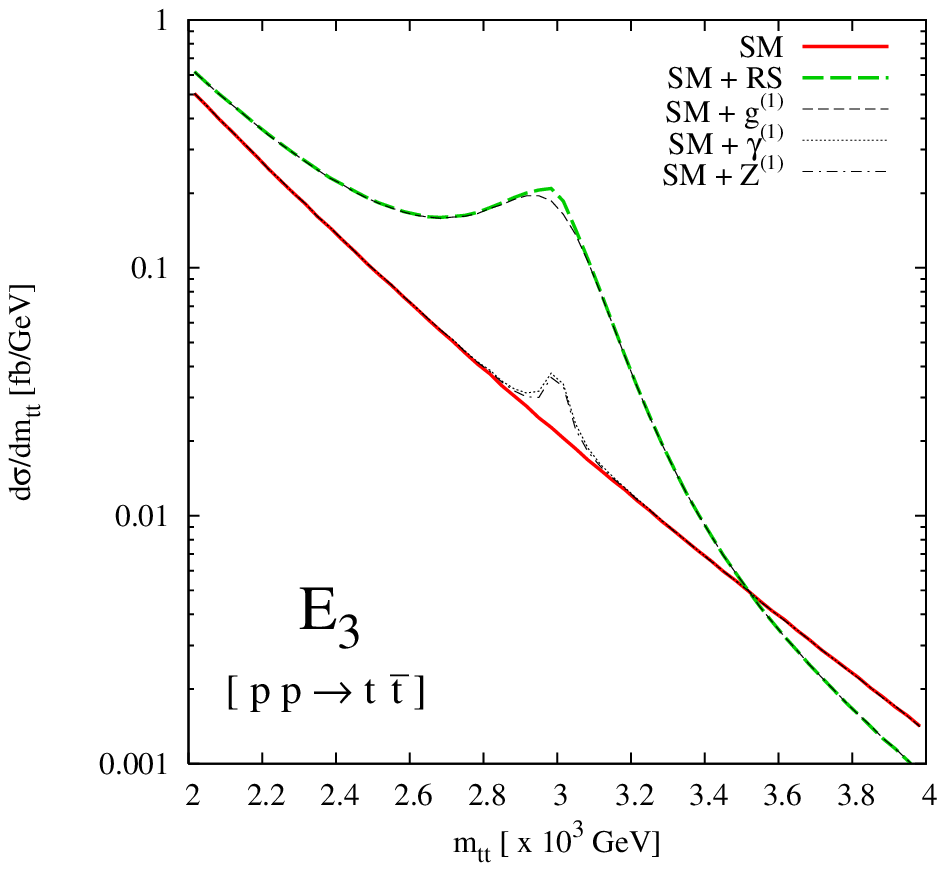,width=8.0cm}
\epsfig{file=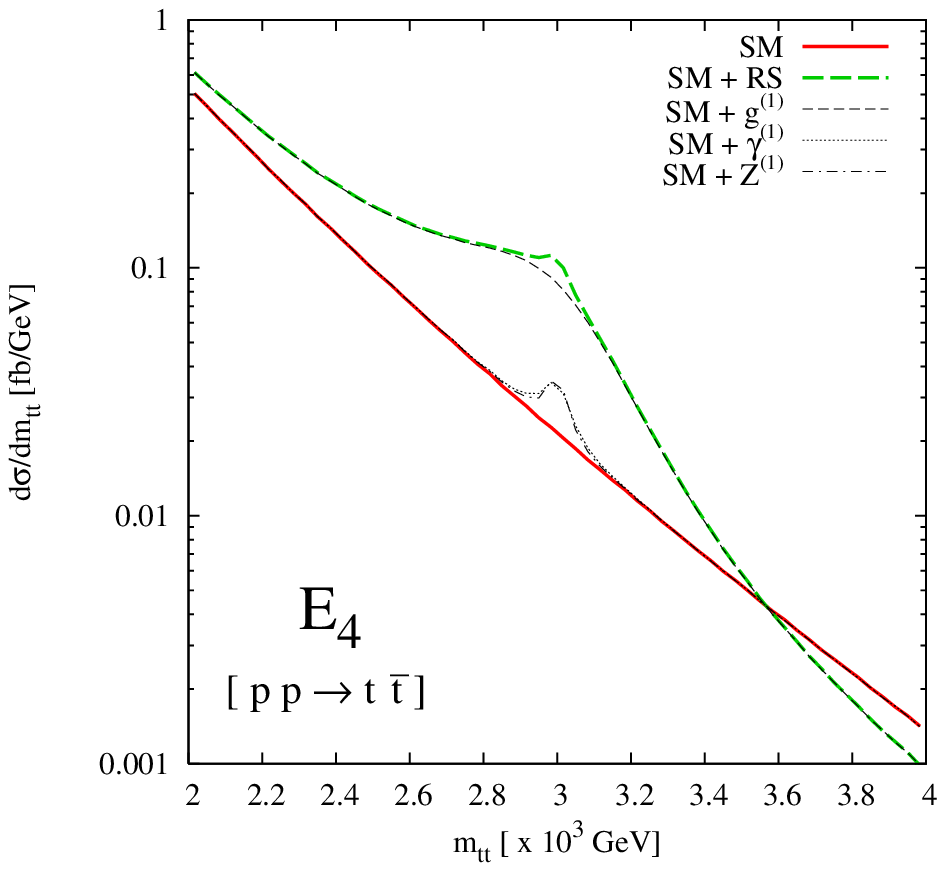,width=8.0cm}
\end{center}
\vspace*{-5mm}
\caption{The invariant mass distribution of the cross section for the 
$pp\to t\bar t$ process for all four scenarios $E_{1,2,3,4}$ defined in 
Table~\ref{tab:par}. We use the CTEQ5M1 set of parton distribution functions 
and include the cuts of  eq.~(\ref{eq:ptcut}) on the $p_T^{t,\bar t}$ and 
$\eta_{t,\bar t}$ of the final quarks. The solid curve is for the SM background
and the long--dashed one for the RS model with contributions of all the first 
KK excitations of the gauge bosons $g^{(1)}, \gamma^{(1)}, Z^{(1)}$ included; 
the short--dashed curve is for the SM case along with only $g^{(1)}$ exchange;
the dotted curve is for the SM case along with only $\gamma^{(1)}$ exchange;
the dashed--dotted curve is for the SM case along with only $Z^{(1)}$ exchange.
}
\label{fig:tt}
%\vspace*{-2mm}
\end{figure}

The invariant mass distributions d$\sigma$/d$m_{t\bar t}$ of the process $pp \to
t\bar t$, are shown in Fig.~\ref{fig:tt} for the four scenarios $E_1$ to $E_4$;
the cuts on the transverse momenta and the rapidity of the final $t$--jets given
in eq.~(\ref{eq:ptcut}) have been applied.  Shown are the distributions in the
case of the SM background, $qq/gg \to t\bar t$, in the case where the KK
excitation of the gluon is also exchanged, $q\bar q \to g,g^{(1)}  \to t\bar
t$,  and in the case where all the KK excitations of the electroweak  gauge
bosons are also exchanged $q\bar q \to g,g^{(1)},\gamma^{(1)},Z^{(1)}, Z^{'(1)}
\to t\bar t$.  As already mentioned, we have used  a common mass for the KK
gauge bosons,  $\mkk =3$ TeV, with the couplings given in Table 1 that lead
to the $b\bar b, t\bar t$ branching ratios and total decay widths displayed in
Table 2.\s 

As can be seen from the figure, there is a substantial contribution of the KK
excitations to the invariant mass distribution, in particular around the peak
$m_{t\bar t} \sim 3$ TeV. At higher invariant masses the KK contribution becomes
 small, while at invariant masses lower than $\mkk$ it is significant even
for $m_{t\bar t} \sim 2$ TeV; only for $m_{t\bar t} \lsim 1$ TeV the KK
contribution becomes negligible. Outside the KK mass peak, the RS effect
is mostly due to the interference between the excited state and SM 
contributions; this interference is positive below and negative above the peak
[where the real part of the propagator of the KK state changes sign].  The
dominant contribution compared to the SM case is by far due to the exchange of
the excitation of the strongly interacting gluon which has the largest (QCD
versus EW) couplings to the initial state partons. The contributions of the KK
photon and KK $Z$ boson increase the peak only slightly. In turn, the KK excitation
of the $Z'$ boson has a negligible impact on $t \bar t$ production 
as it does not couple to the initial light quarks, $Q'(c_{\rm light})=0$,  
and the parton density of the heavier bottom quark in the proton is small.\s

\begin{table}[!h]
\renewcommand{\arraystretch}{1.3}
\begin{center}
\begin{tabular}{|c|cccc|ccc|}\hline
       & SM  & SM+RS & $\ {\cal S}_{10}^{\rm RS}\ $ 
                                       & $\ {\cal S}_{100}^{\rm RS}\ $   
                                       & $\ {\cal S}_{100}^{g^{(1)}}\ $
                                       & $\ {\cal S}_{100}^{\gamma^{(1)}}\ $
                                       & $\ {\cal S}_{100}^{Z^{(1)}}\ $ \\ 
\hline 
$E_1$  & 47.0& 135   & 41  & 128 &124 & 5.7& 7.4 \\ 
$E_2$  & 90.5& 181   & 30  & 95  & 91 & 5.2& 7.0 \\ 
$E_3$  & 16.6& 83.4  & 52  & 164 &157 & 7.8& 7.0\\    
$E_4$  & 34.1& 88.7  & 30  & 94  & 88 & 6.4& 6.4 \\  \hline  
\end{tabular}
\end{center}
\vspace*{-1mm}
\caption{Left: The $pp \to t\bar t$ cross sections [in fb] in the SM and in 
the RS model with the exchange of all the KK excitations as well as the 
significance of the signal for ${\cal L}=10$ fb$^{-1}$ and ${\cal L}=100$ 
fb$^{-1}$. Right: the significance for ${\cal L}=100$ fb$^{-1}$ of the 
signal when only the separate contributions of the KK excitations of the 
gluon, photon and $Z$ bosons are added to the SM contribution. These numbers 
are obtained after the cuts of  eqs.~(\ref{eq:qqcut}) and (\ref{eq:ptcut}) are applied.}
\label{tab:sigtt}
\vspace*{-1mm}
\end{table}

The total cross sections of the $pp \to t\bar t$ process are displayed in
Table \ref{tab:sigtt} first for the SM background only and then in the RS model
in which all contributions of the KK excitations [gluon, photon and $Z$ boson]
are added; the cuts on the invariant mass, transverse momentum and rapidity of
the $t$ and $\bar t$ jets, eqs.~(\ref{eq:qqcut}) and (\ref{eq:ptcut}), are
applied. The significance of the full signal with a common mass, $\mkk=3$ TeV,
is also shown for the integrated luminosities ${\cal L}=10$ and 100 fb$^{-1}$.
As can be seen, the
significance of the excess of events in the RS scenarios when all KK excitations are 
included is large in the four cases, ${\cal S}_{10}^{\rm RS} \gsim 30$ for a 
moderate luminosity and ${\cal S}_{100}^{\rm RS}\gsim 90$ for an high
luminosity.\s  

To illustrate the impact of each KK state separately, the right--hand side of
Tab.~\ref{tab:sigtt} shows the significance of the  signal in the case where
only one KK excitation is exchanged in the $pp\to  t\bar t$ process, that is,
when the mass of the two other excitations is (artificially) set to high values.
As one might have  expected, since the excess over the SM background is mainly
due to the exchange of the KK gluon, the significance in this case, ${\cal
S}_{100}^{g^{(1)}}$, is almost the same as when the full signal is considered,
${\cal S}_{100}^{\rm RS}$.  In the case where only the first KK excitation of
the photon or the $Z$ boson  is considered, the significance  is much smaller. 
This is a mere consequence of the fact that the electroweak couplings of the
$\gamma^{(1)}, Z^{(1)}$ bosons are much smaller than the QCD couplings of the
$g^{(1)}$,   leading to limited production cross sections. The smaller rates
are, however,  partly compensated by the smaller total decay widths and one
obtains in  some scenarios significance at high luminosities that are large
enough  for the effects to be detectable at the LHC, ${\cal
S}_{100}^{\gamma^{(1)}, Z^{(1)}} \gsim 5$.\s

The significance of the signal strongly depends on the chosen scenario 
which fixes the couplings of the KK excitations to the $t,b$ quarks. If the
couplings to top quarks are large, the $pp \to \vkk  \to t\bar t$ production
rate is substantial but the total decay width is also large leading to a more
diluted signal. The signal is even more diluted when the couplings of the KK
excitations to $b$ quarks is at the same time large: on the one hand, this
increases the total decay width of the excitation and, on the other hand, it
lowers the branching ratio into the $t\bar t$ final states that we are looking
at. Taking the example of the $g^{(1)}$ state, the best significance is obtained
in scenario $E_3$ in which the couplings to the $b$ quarks are not too strongly
enhanced (see Table \ref{tab:par}) 
so that the $g^{(1)}$ total decay width is the smallest,
$\Gamma_{g^{(1)}} \simeq 330$ GeV (see Table \ref{tab:br}), 
and the branching ratio for decays into top
quarks is among the largest ones, BR$(g^{(1)} \to t\bar t) \simeq 83\%$. In scenario
$E_{1}$, where the parameters are as in scenario $E_3$ except for the $Q(c_{t_R})$
and $Q'(c_{t_R})$
charges which are larger, the total decay width of $g^{(1)}$ doubles,
$\Gamma_{g^{(1)}} \simeq 630$ GeV, while the fraction for decays into top quarks
is not significantly affected, BR$(g^{(1)} \to t\bar t)\sim 91\%$, leading to a
smaller significance, ${\cal S}_{10}^{\rm RS}=41$ versus ${\cal S}_{10}^{\rm
RS}=52$. In scenarios $E_2$ and $E_4$, the total decay widths of $g^{(1)}$ are
larger and the $t\bar t$ branching ratios smaller so that the significance of
the signals is reduced compared to the two previous scenarios.\s 

\begin{figure}[!h]
\epsfig{file=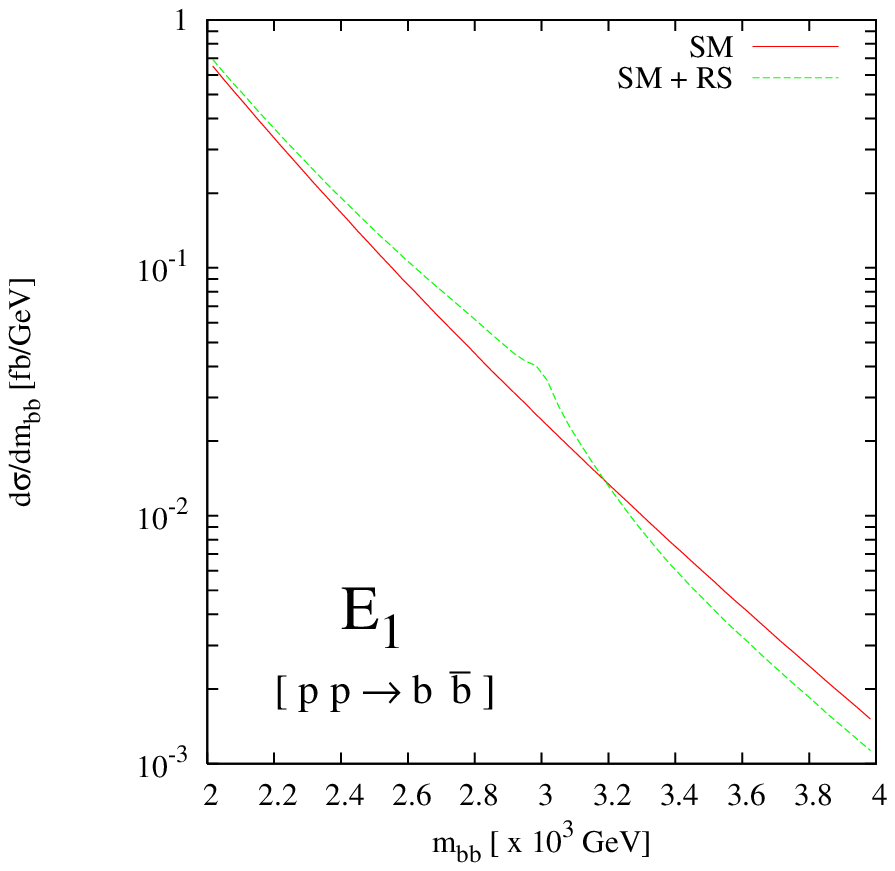,width=8.0cm}
\epsfig{file=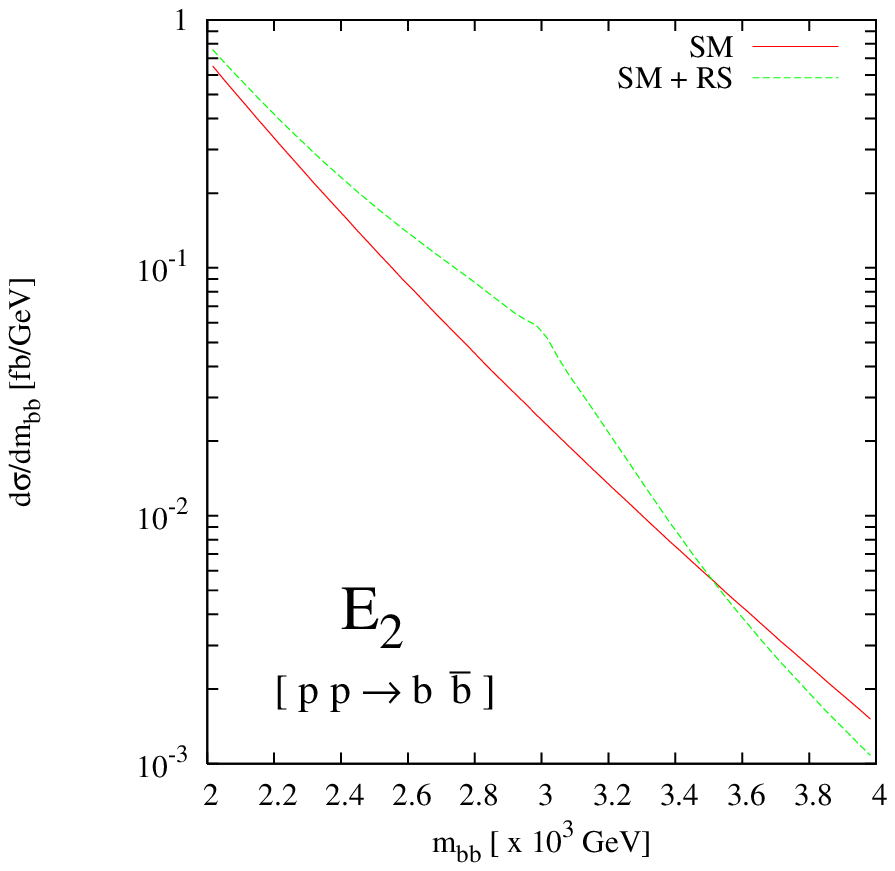,width=8.0cm}\vspace*{5mm}
\epsfig{file=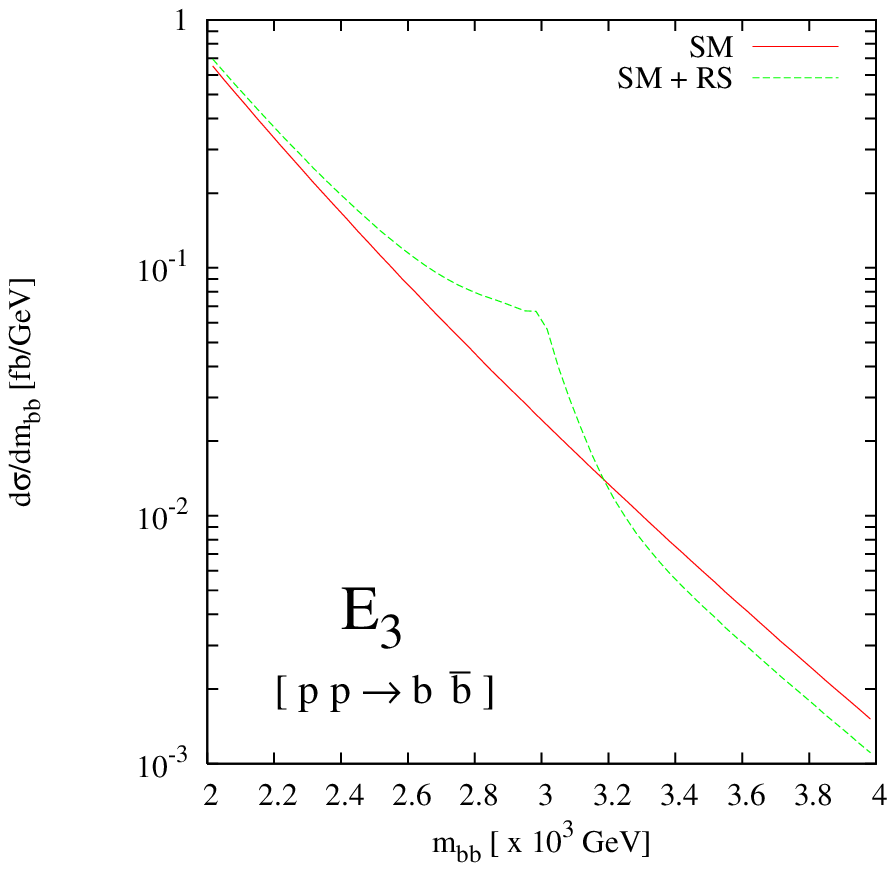,width=8.0cm}\hspace*{3mm}
\epsfig{file=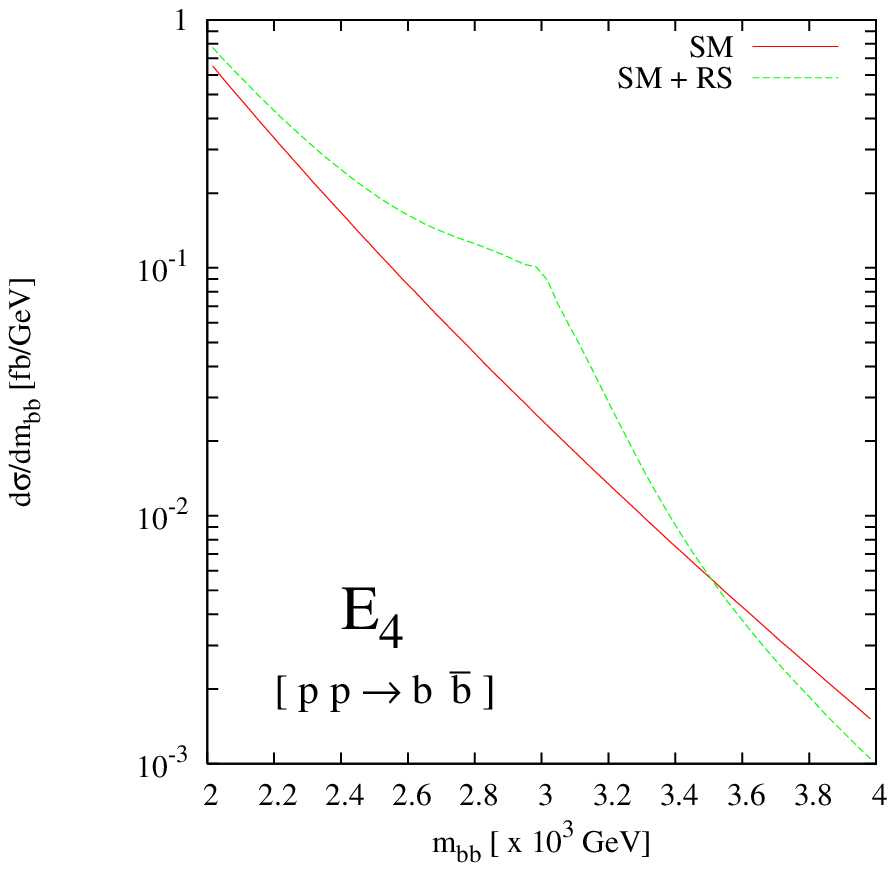,width=8.0cm}
\vspace*{-2mm}
\caption{The invariant mass distribution of the cross section for the 
$pp\to b\bar b$ process for all four parameter sets $E_{1,2,3,4}$ defined in 
Table~\ref{tab:par}. We use the CTEQ5M1 set of parton distribution functions 
and include the cuts of  eq.~(\ref{eq:ptcut}) on the $p_T^{b,\bar b}$ and 
$\eta_{b,\bar b}$ of the final quarks. The solid curve is for the SM background
and the long--dashed one for the RS model with contributions of all the first 
KK excitations of the gauge bosons $g^{(1)}, \gamma^{(1)}, Z^{(1)}, Z^{\prime (1)}$ included.}
\label{fig:bb}
\vspace*{-1mm}
\end{figure}

The discussion for the production process $pp \to b\bar b$ is quite similar to that of
$pp \to t\bar t$ except for the fact that the $t$--channel  contribution  to 
the $b\bar b \to b\bar b$ subprocess has to be added. The  smallness of the
parton distribution function of the bottom quark can be partly compensated  by
the potentially larger KK gauge couplings of the $b_{L,R}$ states  [for small
$c_{b_R}$ values as shown in Table \ref{tab:par}] compared to  light quarks. For
instance, this $b \bar b$ contribution reaches the level of $\sim 10\%$ in the
SM+RS cross section in scenario $E_4$. However, this contribution does not peak
at a given invariant mass value. The invariant mass distribution d$\sigma$/d$
m_{b \bar b}$ for bottom quark pair production is shown in Fig.~\ref{fig:bb} for
$m_{b\bar b}$ between 2 and 4 TeV. As previously, the cuts on the transverse
momenta and rapidity of the final jets given in eq.~(\ref{eq:ptcut}) have been 
applied. Here, we simply show the SM background and the signal excess  in the
case where the contributions of all KK excitations are simultaneously included;
again, this excess is largely dominated by the exchange of the KK excitation of
the gluon.\s

The signals are less striking than in the $pp \to t\bar t$ case, the main reason
being due to the fact that the branching ratios BR($g^{(1)} \to b\bar b)$  are
much smaller than BR($g^{(1)} \to t\bar t)$,
due to the hierarchy between the $b$ and $t$ couplings (see Table \ref{tab:par}),
except in scenario $E_4$ and, to a
lesser extent scenario $E_2$, in which they are comparable. 
The cross section in the SM and in the RS model, as well as the 
significance of the signal for the two possible luminosities  ${\cal L}=10$ and
100 fb$^{-1}$, are shown in Table \ref{tab:sigbb} in the four scenarios when the 
cuts of  eqs.~(\ref{eq:qqcut}) and (\ref{eq:ptcut}) are applied. One can see
that in all cases the significance is large enough, ${\cal S}_{10}^{\rm RS}\gsim
5$, to allow for the detection of the signal even at moderate luminosities. 
For example, in the parameter set $E_4$, the $c_{b_R}$ value is smaller 
than for $E_3$ so that
the KK gauge coupling to $b$ is larger, leading to higher $b \bar b$ production cross
section and significance.
Comparing now $E_4$ to $E_2$, $c_{t_R}$ is larger which induces smaller widths for KK gauge
bosons and thus a better significance, for same reasons as before.\s

\begin{table}[!h]
\renewcommand{\arraystretch}{1.3}
\begin{center}
\begin{tabular}{|c|cccc|}\hline
       & SM  & SM+RS & $\ {\cal S}_{10}^{\rm RS}\ $ 
                       & $\ {\cal S}_{100}^{\rm RS}\ $ \\ \hline 
$E_1$  & 55.3& 67.6  & 5.2 & 17   \\ 
$E_2$  & 109 & 159   & 15  & 48    \\ 
$E_3$  & 17.0& 32.8  & 12  & 38   \\    
$E_4$  & 39.7& 90.0  & 25  & 80   \\  \hline  
\end{tabular}
\end{center}
\vspace*{-1mm}
\caption{The $pp \to b\bar b$ cross sections [in fb] in the SM and in 
the RS model with the exchange of all the KK excitations as well as the 
significance of the signal for ${\cal L}=10$ fb$^{-1}$ and ${\cal L}=100$ 
fb$^{-1}$; the cuts of  eqs.~(\ref{eq:qqcut}) and (\ref{eq:ptcut}) have been
applied.}
\label{tab:sigbb}
\vspace*{-1mm}
\end{table}

Note that when combining the $pp \to t\bar t$ and $pp \to b\bar b$ processes,
one  would in principle be able to have access to the couplings of the KK 
states $g^{(1)}$ to top and bottom quarks. However, it would have only a small
contamination from the various overlapping electroweak KK resonances, since the
major part of the signal is due to the contribution of the gluon excitation. 
As the couplings are parity violating,  an additional and  important
information is provided by some polarization and forward--backward asymmetries
to which we turn our attention now. 

\subsubsection*{3.3 Polarization and forward--backward asymmetries}
\label{sub:polarizations}

Each of the KK excitations of the gauge bosons, $g^{(1)}, \gamma^{(1)}$ and
$Z^{(1)}$, has a different coupling to the right-- and left--handed top quarks 
[which are themselves different from the SM ones]. These couplings  appear in
the forward--backward asymmetry as well as in the polarization of the  produced
top quarks. While the enhancement in the production rate due to a single KK
gauge boson is proportional to the sum of squared couplings $(g_{Q_L}^{\vkk })^2
+(g_{Q_R}^{\vkk })^2$,  the polarization and forward--backward asymmetries are
proportional to the difference $(g_{Q_L}^{\vkk })^2 -(g_{Q_R}^{\vkk })^2$. Thus,
a combined measurement of the cross section together with asymmetries would
determine the couplings of  the vector boson $\vkk $. However, since the process
is mediated by the exchange of several KK gauge bosons with differing right--
and left--handed couplings, it  will not be possible to measure these
couplings for any of the electroweak KK  excitations. 
This is particularly true as the
major contribution to the total rate for $\sigma(pp \to Q\bar Q)$ is coming from
$g^{(1)}$, as the contribution from the electroweak excitations
$\gamma^{(1)}$, $Z^{(1)}$ and $Z^{\prime (1)}$ 
is relatively small. 
Nevertheless, the measurement of the
polarization and forward--backward asymmetries for top quarks, on which we will
concentrate here,  will be instrumental in establishing the presence of parity
violating KK gauge bosons.\s

\begin{figure}[!h]
\begin{center}
\epsfig{file=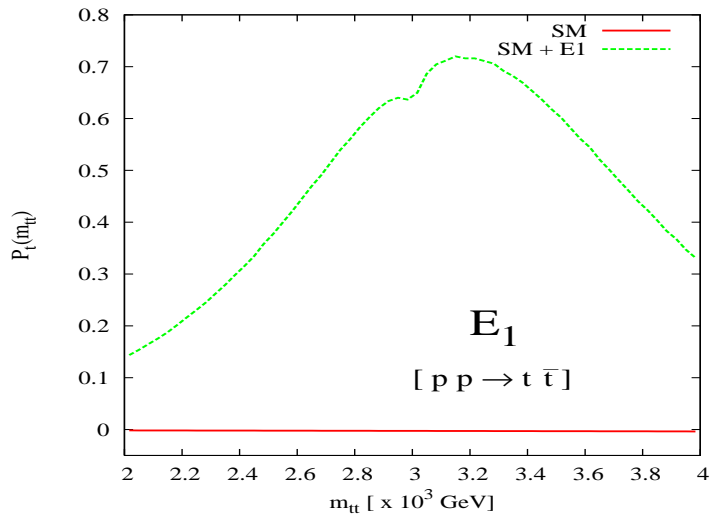,width=9.9cm,height=6.9cm}
\end{center}
\vspace*{-4mm}
\caption{The top quark polarization as a function of the invariant mass $m_{t
\bar t}$ in the reaction $pp \to t \bar t$ in the SM (solid) and in the RS 
model (dashed) for the set $E_1$ of parameters.} 
\label{fig:Ptmtt}
\vspace*{-2mm}
\end{figure}

The polarization of the produced top quarks is defined as:
\begin{equation}
P_t = \frac{N_R - N_L} {N_R + N_L},
\end{equation}
where $N_R$ and $N_L$ are the numbers of events with right-- and left--handed 
helicity for the top quark with a given set of kinematical cuts\footnote{We note
here that even for extra--dimensional models in which the KK excitations have
the same couplings as their SM counterparts, the $Z^{(1)}$ state will always
have parity violating couplings and hence, the produced top quarks will be
polarized. However,  in this case, following the structure of the SM, the left
chiral top quark has larger coupling to the $Z^{(1)}$ excitation, leading to a
negative polarization. On the other hand, in the RS model that we consider here,
the right chiral top quarks have larger couplings to the KK gauge bosons,
leading to a positive polarization.}. The statistical error in the measurement
of this polarization, in terms of the rate $\sigma$ and the LHC luminosity
${\cal L}$, is given by 
\begin{equation} \Delta P_t = \frac{1}{\sqrt{N_R+N_L}} \sqrt{1-P_t^2} 
= \frac{1}{\sqrt{\sigma \times {\cal L}}} \sqrt{1-P_t^2},
\end{equation} 
The polarization of the top quark is shown in Fig.~\ref{fig:Ptmtt} in the case
of the SM and for the RS scenario $E_1$ as a function of the invariant mass 
$m_{t\bar t}$; again the cuts of eqs.~(\ref{eq:ptcut}) have been applied and the
contributions of all virtual KK states has been included.\s 
%The degree of
%polarization in both the SM and RS cases and the significance for the two LHC
%luminosities ${\cal L}=10$ fb$^{-1}$ and ${\cal L}=100$  fb$^{-1}$ in the
%scenarios $E_{1,..,4}$ are displayed in Table \ref{tab:asy}.\s

The polarization is a parity odd quantity and can assume a non--zero value only
in the presence of parity violating interactions. In the SM case, the only 
source of parity violation is the contribution of the $Z$ boson, which leads to
a  small negative polarization of order $\sim 10^{-3}$. The dominant 
contribution for $m_{t\bar t}$ near $M_{KK}$ is due to $g^{(1)}$ which 
has a larger coupling to $t_R$ than to $t_L$ in the chosen RS scenarios; 
this leads to a large polarization roughly of the size 
\beq
  \frac{(g_{t_R}^{\vkk })^2-(g_{t_L}^{\vkk })^2}{ 
(g_{t_R}^{\vkk })^2 +(g_{t_L}^{\vkk })^2} \equiv G_{-+}^{\vkk }
\eeq
coming from the $q\bar q$ fusion channel. However, there is a dilution of this
polarization due to the parity--conserving SM production of top pairs through
$gg$ fusion and  $q\bar q$ annihilation with gluon exchange. For instance, in 
the case of scenario $E_1$, one has for the first gluon excitation $g^{(1)}$, 
$G_{-+}^{g^{(1)}}= 0.86$, while the degree of polarization, $P_t=0.505$, is diluted 
by the SM QCD processes. Similar results are obtained in the other scenarios and
one has: $P_t \simeq 0.37,0.52,0.38$ in scenarios $E_2, E_3,E_4$, respectively. 
Furthermore, there is some subleading contribution from the $\gamma^{(1)}$
and $Z^{(1)}$ states which is, for instance, visible as a small ``dip''
just before the peak  in the $m_{t\bar t}$ distribution of the polarization
$P_t$ in  Fig.~\ref{fig:Ptmtt}.\s 

The polarization of the top quark is not a direct observable.  Nevertheless, it
is reflected for instance in the angular distribution of the lepton coming from
the decaying top quark, $t \to bW \to b\ell \nu$ with  $\ell=e, \mu$. In the
rest frame of the $t$--quark, the leptons issued from these decays are
isotropically distributed for unpolarized top quarks. For positive  $P_t$, the
lepton distribution is peaked in the direction of the ``would be'' boost or the
direction of quantization axis, and for negatively polarized top quarks, the
peak is in the opposite direction. In the laboratory frame, the decay products
of the fast moving top quark are collimated in the forward direction due to the
boost. Thus, for the positively polarized top quarks, there will be additional
focusing of the leptons in the direction of the top quark momenta. Accordingly,
there is a de--focusing of the lepton distribution for the negatively polarized
top quarks. Thus, the simplest quantity to look at is the azimuthal distribution
(or the focusing) of the leptons with respect to the top production plane
\cite{topolar}. We thus define the following  asymmetry in the laboratory 
frame:
\begin{equation}
A_\ell = \frac{\sigma(\cos\phi_\ell>0)-\sigma(\cos\phi_\ell<0)}
{\sigma(\cos\phi_\ell>0)+\sigma(\cos\phi_\ell<0)},
\end{equation}
where $\phi_\ell$ is the azimuthal angle of the decay leptons with respect to 
the $t$--quark production plane, or with respect to the transverse momentum of 
the top quark. 
%This asymmetry is large even in the SM as a result of the large 
%boost of the top quark in the laboratory frame and is, therefore, best suited 
%to measure the negative polarizations. 
Again, the statistical error in the measurement of the asymmetry and the
corresponding significance is given by
\begin{equation}
\Delta A_\ell =  \frac{1}{\sqrt{\sigma \times {\cal L}}} \sqrt{1-A_\ell^2}
\hspace{1.0cm} {\rm and} \hspace{1.0cm} {\cal S} = \frac{A_\ell^{RS+SM} -
A_\ell^{SM}}{\Delta A_\ell^{SM}}
\label{eq:Al}
\end{equation}
respectively.
The values of the asymmetries in the various RS scenarios and the significance 
in their measurements are shown in the right--hand side of Table \ref{tab:asy}. 
%Again, the usual cuts have been applied and all channels have been added.\s

\begin{figure}[!h]
\begin{center}
\epsfig{file=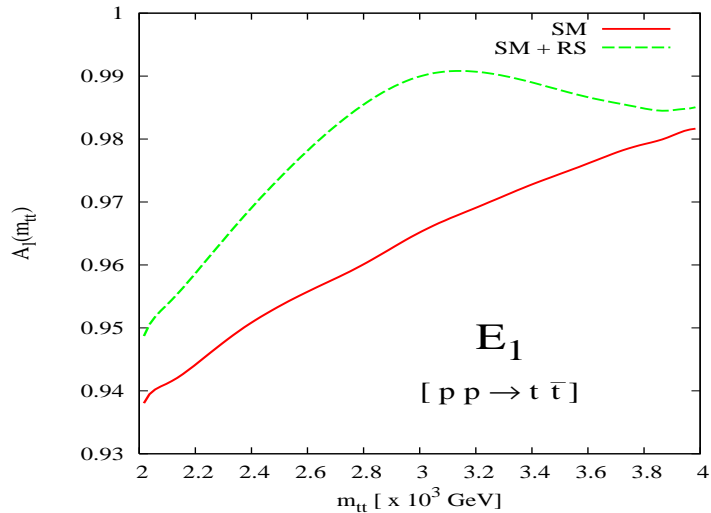,width=9.9cm,height=6.9cm}
\vspace*{-1mm}
\caption{The lepton asymmetry distribution versus the $m_{t\bar t}$ invariant
mass for the process $pp \to t \bar t$ in SM and in the RS scenario $E_1$.
\protect\label{fig:Almtt}}
\end{center}
\vspace*{-2mm}
\end{figure}

Since we are dealing with  heavy KK bosons, the top quarks are highly boosted in
the center of mass frame, which leads to highly collimated  decay products even
for unpolarized top quarks. This explains the large  value of the asymmetry,
$A_\ell \simeq 0.96$, for the SM in the case of $E_1$ cuts for example; see
Table \ref{tab:asy}. A modification of the top polarization changes the
collimation of the decay products and, thus, affects this SM asymmetry. For
instance, again in scenario $E_1$, one expects a large and positive polarization
for the top quarks, which leads to a slightly larger asymmetry value, $A_\ell
\simeq 0.98$. The significance ${\cal S}$ calculated according to
eq.~(\ref{eq:Al}), indicates that this asymmetry  should be observable at 
the LHC but only for a high luminosity run in a post--discovery analysis [the
discovery being clearly possible at a low luminosity] as shown in
Table \ref{tab:asy}. Therefore, the associated distribution displayed  in
Fig.~\ref{fig:Almtt} for the parameter set $E_1$, which exhibits a specific
shape, should help in identifying the RS scenario.  In particular, the fact that
this $A_\ell$ asymmetry is enhanced by the RS contribution proves that the
exchanged KK gauge bosons couple effectively more to the $t_R$ than to the $t_L$
state.\s

\begin{table}[!h]
\vspace*{5mm}
\renewcommand{\arraystretch}{1.5}
\begin{center}
\begin{tabular}{|c|cccc|cccc|}\hline
       & $A_\ell^{\rm SM}$  & $A_\ell^{\rm RS+SM}$ & $\ {\cal S}_{10}^{A_\ell}\ $ 
                                       & $\ {\cal S}_{100}^{A_\ell}\ $   
       & $A_{\rm FB}^{\rm SM}$  & $A_{\rm FB}^{\rm RS+SM}$ & $\ {\cal S}_{10}^{A_
       {\rm FB} }\ $  & $\ {\cal S}_{100}^{A_{\rm FB} }\ $   \\ \hline 
$E_1$  &0.957&0.982&1.5 &4.7 & -0.00178& 0.195 &7.2& 23 \\ 
$E_2$  &0.952&0.972&1.3 &4.1 &  0.00737& 0.157 &6.7& 21 \\ 
$E_3$  &0.963&0.987&1.2 &3.8 & -0.00666& 0.232 &6.7& 21 \\    
$E_4$  &0.960&0.980&1.0 &3.2 & -0.00407& 0.181 &5.4& 17 \\  \hline  
\end{tabular}
\end{center}
\vspace*{-3mm}
\caption{Left: The polarization and forward--backward asymmetries, $A_\ell$ and
$A_{\rm FB}^t$, for  the process $pp \to t\bar t$ in the SM and in 
the RS model with the exchange of all the KK excitations as well as the 
significance of their measurements for ${\cal L}=10$ fb$^{-1}$ and ${\cal L}
=100$ fb$^{-1}$.}
\label{tab:asy}
\vspace*{-3mm}
\end{table}

Let us now turn our attention to the forward--backward asymmetry.  The
$s$--channel diagrams exchanging gauge bosons with parity violating  couplings
lead to a forward--backward asymmetry in the top quark pair production in the
center of mass frame with respect to the quark direction. However, since at the
LHC there are two identical protons colliding, there is no simple
forward--backward asymmetry in the laboratory frame due to the symmetric
distribution of the initial state quarks and anti-quarks. But, since the proton
is mainly composed of quarks and with only a small proportion of anti-quarks in
the sea, the fusing $q\bar q$ pairs are highly boosted in the laboratory frame,
mainly in the direction of the incoming quarks. Thus, one can construct a kind
of forward--backward asymmetry with respect to the direction of the boost in the
center of mass frame. In the laboratory frame, we first arbitrarily label the
proton beams as ``1'' and ``2'', and then  define a quantity
\beq
C = \left(\frac{x_1-x_2}{x_1+x_2}\right) \cos\theta_t^{\rm lab}
\eeq
where $x_i$ are the momenta fractions of the colliding partons and $\theta_t^{
\rm lab}$  the $t$--quark polar angle with respect to the beam ``1'';
%Note that $C$ is even under relabeling of the proton beams and hence can 
%assume a non--zero value. 
the quantity in the brackets is the boost of the center of mass frame in the 
laboratory frame. With respect to this quantity we define a forward--backward 
asymmetry with  respect to the direction of boost: 
\begin{equation}
A_{\rm FB}^t = \frac{\sigma(C>0)-\sigma(C<0)}{\sigma(C>0)+\sigma(C<0)}.
\end{equation}
This quantity is closely related to the forward--backward asymmetry in the 
center of mass frame, with respect to the quark direction, up to some smearing
due to the  boost of the center of mass frame and possible misidentification of
a $\bar q$  as a $q$ for a faster moving $\bar q$. The statistical error in the
measurement of this asymmetry is given by
\begin{equation}
\Delta A_{\rm FB}^t =  \frac{1}{\sqrt{\sigma \times {\cal L}}} \sqrt{1-(A_{\rm 
FB}^t)^2}.
\end{equation}

\begin{figure}[t]
\vspace*{2mm}
\begin{center}
\epsfig{file=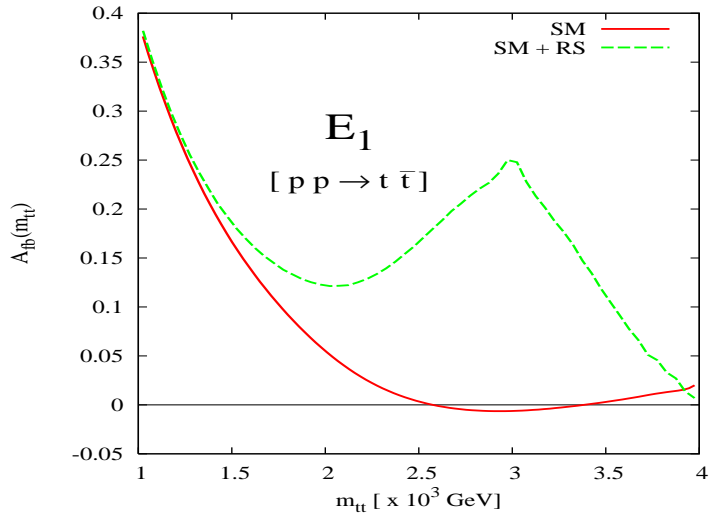,height=6.9cm,width=9.9cm}
\caption{The forward--backward asymmetry distribution versus the invariant mass
$m_{t\bar t}$ for the reaction $pp \to t \bar t$ in the SM and in the RS 
scenario $E_1$.
\protect\label{fig:Afbmtt}}
\end{center}
\vspace*{-3mm}
\end{figure}

We should note that the above asymmetry is not a pure probe of parity violation,
i.e. it can be non--zero and positive even for parity--conserving processes.  The
reason is the boost of the center of mass frame with respect to the quark 
direction, which leads to focusing of particles in the forward direction. This 
explains the large value of $A_{\rm FB}^t$ for the SM as shown in
Fig.~\ref{fig:Afbmtt} by a solid line. Additional sources of a forward--backward
asymmetry in the center of mass frame change the value of $A_{\rm FB}^t$, which
is shown in Fig.~\ref{fig:Afbmtt} by a dashed line for the RS scenario $E_1$
where there is a significant  contribution from parity violating couplings of
the $g^{(1)}$ excitation. Once  more, there is some finite contribution from the
electroweak  KK modes which  appears as a ``nipple'' peak in the $m_{t\bar t}$
distribution of  $A_{\rm FB}^t$, Fig.~\ref{fig:Afbmtt}.\s 

The obtained values of $A_{\rm FB}^t$ and their  significance  given in
Table \ref{tab:asy} for all scenarios, shows that it is potentially visible at
the LHC even at low luminosity with $\ {\cal S}_{10}^{A_{\rm FB} }\ \ge 5$. 
Such significant shifts in the asymmetry with respect to the SM expectation
would strongly indicate parity violation in the top couplings to KK gauge
bosons. Indeed, these typical asymmetry deviations cannot be entirely explained
by interferences between hypothetical parity--conserving bosons and the parity
violating SM $Z$ boson. Moreover, the profile of the $A_{\rm FB}^t$ distribution
in Fig.~\ref{fig:Afbmtt} is characteristic of the exchange of several KK gauge
bosons. The enhancement of $A_{\rm FB}^t$ originates from the fact that the
effective couplings of KK gauge bosons, and mainly the KK 
gluon $g^{(1)}$, are larger for the $t_R$ than for $t_L$.\s

Note that we have also calculated the asymmetry $A_{\rm FB}^b$ for the $ pp \to
b\bar b$ production process, but we do not present the corresponding  results as
the significance is rather low in all the considered scenarios.  Furthermore, we
will not discuss the polarization of the produced bottom quarks as it will not
be experimentally measurable.

\subsection*{4. The higher order processes}

\subsubsection*{4.1 The gluon--gluon fusion mechanism}
\label{sub:associated}

\begin{figure}[!h]
\begin{center}
\vspace*{0.8cm}
\hspace*{-3cm}
\begin{picture}(300,50)(0,0)
\SetWidth{1.1}
%\SetScale{1.1}
\Gluon(0,0)(50,0){4}{5}
\Gluon(0,50)(50,50){4}{5}
\ArrowLine(90,25)(50,50)
\ArrowLine(50,50)(50,0)
\ArrowLine(50,0)(90,25)
\Photon(90,25)(140,25){4}{5}
\ArrowLine(140,25)(170,35)
\ArrowLine(140,25)(170,15)
\Text(-20,60)[]{(a)}
\Text(65,25)[]{$Q'$}
\Text(110,40)[]{$\vkk$}
\Text(180,40)[]{$Q$}
\Text(180,10)[]{$\bar Q$}
\Text(0,40)[]{$g$}
\Text(0,10)[]{$g$}
\hspace*{2cm}
\SetOffset(175,0)
\Gluon(0,50)(40,25){4}{5}
\Gluon(0,0)(40,25){4}{5}
\DashLine(40,25)(80,25){4}
\Photon(80,25)(120,25){4}{5}
\ArrowLine(120,25)(150,40)
\ArrowLine(120,25)(150,10)
\Text(-20,60)[]{(b)}
\Text(100,37)[]{$\vkk$}
\Text(65,35)[]{$V_5$}
\Text(158,10)[]{$Q$}
\Text(158,40)[]{$\bar Q$}
\Text(0,35)[]{$g$}
\Text(0,15)[]{$g$}
\end{picture}
\end{center}
\caption[]{\it The Feynman diagram for $Q \bar Q$ ($Q=b,t$) 
production, via $\vkk$ exchange, at the LHC through 
the quark loop induced gluon-gluon fusion process 
(a) and the counter-term with the St\"{u}ckelberg mixing that regulates the 
gauge anomaly of the $gg\vkk$ vertex (b).}
\protect\label{fig:ggV} 
\vspace*{1mm}
\end{figure}
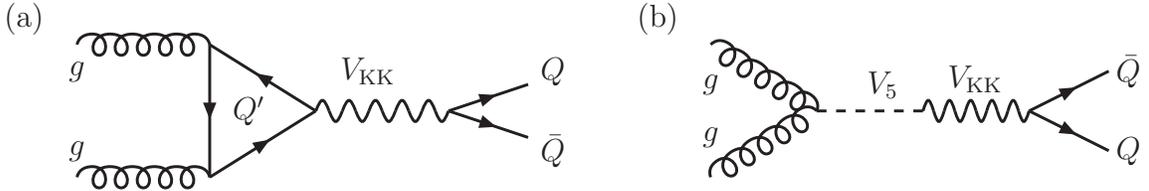

The higher order gluon--gluon fusion contribution to the heavy quark  pair
production at LHC, via KK gauge boson exchange, should be  favored by the large
$V_{KK}$ couplings to heavy quarks. Such a reaction occurs through a loop
involving mainly the heavy top and bottom quarks  and a $V_{KK}$ exchanged in
the s-channel, as shown in Fig.~\ref{fig:ggV}(a).  Indeed, the main reaction is
via $V_{KK}=g^{(1)}$ which receives no   contributions from the light quarks in
the loop as those have quasi identical $g^{(1)}$ couplings for the left and
right chirality (fixed by a common $Q^{(\prime)}(c_{light})$ value as explained
at the end of Section 2.2) which makes the final amplitude vanishing.  
This reaction represents the dominant two--body
production mechanism for KK  excitations which do not couple to light quarks.\s

The effective one--loop vertex $g(k_1^\nu)g(k_2^\rho)V_{KK}(k_3^\mu)$, where
$k_1^\nu,k_2^\rho,k_3^\mu$ denote the respective momenta ($k_1$ and $k_2$ being
taken as incoming with respect to the flow in the loop), is given in momentum
space by, % ,
\begin{equation}
\Gamma^{ijk}_{\mu \nu \rho} = 
t^{ijk} \bigg [
\bigg ( A_1(k_1,k_2) - B_2(k_1,k_2) \bigg )
\epsilon_{\mu \nu \rho \sigma}
(k_1^\sigma-k_2^\sigma)
+ \frac{B_1(k_1,k_2)}{2 k_1 . k_2 }
k_2^\sigma k_1^\tau 
(k_2^\nu \epsilon_{\mu \rho \sigma \tau}
-k_1^\rho \epsilon_{\mu \nu \sigma \tau})
\bigg ],
\label{eq:LoopAmp}
\end{equation}
with,
\begin{equation}
t^{ijk} = g_s^2
Tr [ \lambda_i \lambda_j \lambda_k ]
\sum_f \bigg [
g_{q_{L}}^{V} Q^{(\prime)} (c_{q_{L}})
-
g_{q_{R}}^{V} Q^{(\prime)} (c_{q_{R}})
\bigg ],
\label{eq:trace}
\end{equation}
$g_s$ being the strong interaction coupling constant, 
$\lambda_i$ $[i=1,\dots,8]$ the Gell-Mann matrices (only two such
matrices must be taken in the case of colorless KK excitations: $V_{KK}$ 
$\neq$ $g^{(1)}$), $\epsilon_{\mu \nu \rho \sigma}$ 
the purely antisymmetric tensor and $q_{L/R}$ the quarks running in the loop.
Finally, the $A_1(k_1,k_2)$ function, entering eq.~(\ref{eq:LoopAmp}), 
represents the usual amplitude a priori divergent.
The finite functions $B_{1,2}(k_1,k_2)$  are  defined later.\s

Indeed, the amplitude eq.~(\ref{eq:LoopAmp}) contains an anomaly in the third current. 
As a matter of fact, only two of the three Ward identities that the vertex must obey 
are satisfied\footnote{See Ref.~\cite{Binetruy,Hamed} for the generic 
computations of gauge
anomalies in the context of $S^{1}/\mathbb{Z}_{2}$ orbifold models.}:
$$
k_1^\nu \Gamma^{ijk}_{\mu \nu \rho}=0, 
\ \ \
k_2^\rho \Gamma^{ijk}_{\mu \nu \rho}=0, 
$$
\begin{equation}
-k_3^\mu \Gamma^{ijk}_{\mu \nu \rho}=
-(k_1^\mu+k_2^\mu)\Gamma^{ijk}_{\mu \nu \rho}=
- t^{ijk} 2 (A_1(k_1,k_2) - B_2(k_1,k_2)) \ \epsilon_{\nu \rho \sigma \tau} k_2^\sigma k_1^\tau .
\label{Anomaly}
\end{equation}
The first two relations allow to define the ambiguous amplitude $A_1(k_1,k_2)$ in terms
of finite functions:
\begin{equation}
A_1(k_1,k_2)-B_2(k_1,k_2)-\frac{B_1(k_1,k_2)}{2}=0.
\label{eq:Aexpress}
\end{equation}

However, the $ggV_{KK}$ vertex receives another contribution from
the exchange of the gauge field fifth component $V_5$, 
as drawn in Fig.~\ref{fig:ggV}(b).
There, the $ggV_5$ coupling originates from the so-called 
Chern--Simons term in the action 
(see Ref.~\cite{Hirayama} for the case of warped orbifolds) 
which can be written in terms of the gauge
covariant field strengths as \cite{Hill,Serone}:
\begin{equation}
S_{CS} = 
\int d^4x dy \rho(y) \epsilon^{MNPQR} 
Tr \bigg [ 
A_M F_{NP} F_{QR} 
+ i A_M A_N A_P F_{QR} 
-\frac{2}{5} A_M A_N A_P A_Q A_R
\bigg ],
\label{CSterm}
\end{equation}
where the Roman letter indices $M,N$\dots run over all dimensions.
While the mixing between $\partial_\mu A_5^{(n)}$ and $A_\mu^{(n)}$
(the Greek letter index $\mu$ 
is running over the first four dimensions only), 
appearing in the diagram of Fig.~\ref{fig:ggV}(b), and proportional
to the gauge KK mass, comes from the St\"{u}ckelberg term; see for
instance Refs.~\cite{Hill,Dudas}.
Following Ref.~\cite{Dudas}, 
one finds that the diagram of Fig.~\ref{fig:ggV}(b)
gives rise to a vertex contribution of the form,
\begin{equation}
\Gamma^{ijk}_{\mu \nu \rho}\vert_{CS} = 2 t^{ijk}
A_1(k_1,k_2) \frac{k_{1\mu}+k_{2\mu}}{(k_1+k_2)^2} 
\epsilon_{\nu \rho \sigma \tau} k_2^\sigma k_1^\tau .
\label{CScontrib}
\end{equation}
The whole amplitude is then free from any gauge anomaly as one has now, 
$$
k_1^\nu(\Gamma^{ijk}_{\mu \nu \rho}+\Gamma^{ijk}_{\mu \nu \rho}\vert_{CS})=0, 
\ \ \
k_2^\rho(\Gamma^{ijk}_{\mu \nu \rho}+\Gamma^{ijk}_{\mu \nu \rho}\vert_{CS})=0, 
$$
\begin{equation}
- k_3^\mu(\Gamma^{ijk}_{\mu \nu \rho}+\Gamma^{ijk}_{\mu \nu \rho}\vert_{CS})=
2 t^{ijk} B_2(k_1,k_2)  \epsilon_{\nu \rho \sigma \tau} k_2^\sigma k_1^\tau ,
\label{AnomCancel}
\end{equation}
where $B_2$ is proportional to the mass of fermions running in the loop.
Therefore, as expected, there is a mechanism responsible for the cancellation
of the gauge anomaly.\s

The $B$ functions entering above equations read as,
$$
B_1(k_1,k_2) =
\frac{2}{3 (2\pi)^2} \int_0^1 dz \int_0^{1-z} dx \
\frac{x(x+z-1)}{(m_f^2/(2 k_1.k_2))- x z},
$$
\begin{equation}
B_2(k_1,k_2) =
\frac{2}{3 (2\pi)^2} \int_0^1 dz \int_0^{1-z} dx \
\frac{(1-z) m_f^2}{ m_f^2-(2 k_1.k_2) x z},
\label{eq:B}
\end{equation}\s

In conclusion, both the vertex contributions eq.~(\ref{eq:LoopAmp})  and
eq.~(\ref{CScontrib}) have to be taken into account in the calculation of the
quark contribution to the  $gg\vkk$ vertex. Note that the correction
eq.~(\ref{CScontrib}) vanishes in the on-shell regime limit for $V_{KK}$ where
$k_3^\mu \epsilon_{3\mu}=0$, $\epsilon_3$ being its polarization vector. This is
consistent with Yang's theorem according to which the entire production cross
section vanishes if the two initial zero-mode gluons as well as the produced KK
gauge boson are simultaneously on-shell. This is illustrated by the
$m_{t\bar t}$ invariant mass distribution that we will  discuss now.\s

\begin{figure}[!h]
\centerline{\epsfig{file=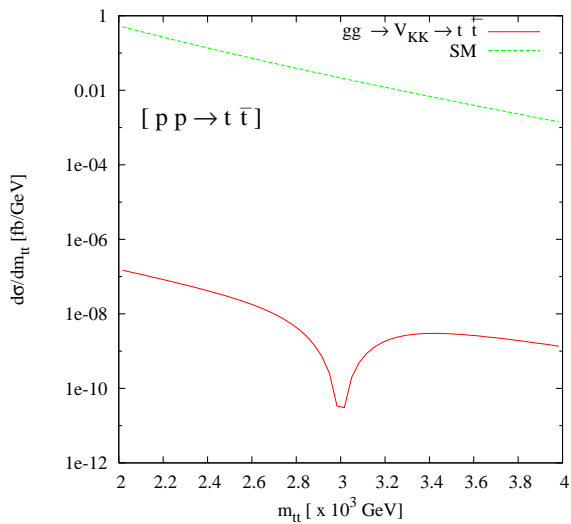,height=7.0cm}}
\caption[]{\it The $m_{t\bar t}$ distribution of the cross-section
corresponding to the reaction shown in Fig.~\ref{fig:ggV} for 
$\vkk = g^{(1)}$
for the $E_2$ choice of parameters. The whole SM distribution
is also shown for comparison.}
\protect\label{fig:ggV-mtt}
\end{figure}

The $m_{t\bar t}$ distribution for the above said process is shown in
Fig.~\ref{fig:ggV-mtt}, along with that for the SM background, for the 
excitation $g^{(1)}$ and the $E_2$ choice of RS parameters. The dip in the
distribution near the mass $m_{t\bar t}=m_{KK}$ is due to the vanishing coupling
for on--shell $\vkk$ bosons with a pair of gluons. The shape of the distribution
depends  upon the total decay width of the $\vkk$. However, the rates for the 
signals are several orders of magnitude smaller than the SM background. The
cross sections are even smaller for the KK excitations of the electroweak
gauge bosons. One can then conclude that the gluon--gluon fusion mechanism
for the production of KK excitations will not be relevant at the LHC. 

\subsubsection*{4.2 Associated production with heavy quarks}

\begin{figure}[!h]
\begin{center}
\vspace*{0.8cm}
\begin{picture}(300,50)(0,0)
\SetWidth{1.1}
%\SetScale{1.1}
\Gluon(0,0)(50,0){4}{5}
\Gluon(0,50)(50,50){4}{5}
\ArrowLine(50,50)(100,60)
\Line(50,50)(50,0)
\ArrowLine(50,0)(100,-10)
\Photon(50,25)(90,25){4}{5}
\ArrowLine(90,25)(120,35)
\ArrowLine(90,25)(120,15)
\Text(-20,60)[]{(a)}
\Text(109,60)[]{$Q'$}
\Text(109,-10)[]{$\bar Q'$}
\Text(70,39)[]{$\vkk$}
\Text(130,40)[]{$Q$}
\Text(130,10)[]{$\bar Q$}
\Text(0,40)[]{$g$}
\Text(0,10)[]{$g$}
\hspace*{1cm}
\SetOffset(175,0)
\Gluon(0,0)(50,0){4}{5}
\ArrowLine(0,50)(50,50)
\ArrowLine(50,50)(50,0)
\ArrowLine(50,0)(100,-10)
\Photon(50,50)(100,50){4}{5}
\ArrowLine(100,50)(130,60)
\ArrowLine(100,50)(130,40)
\Text(-20,60)[]{(b)}
\Text(139,60)[]{$Q$}
\Text(70,35)[]{$\vkk$}
\Text(105,-10)[]{$b$}
\Text(139,40)[]{$\bar Q$}
\Text(0,40)[]{$b$}
\Text(0,10)[]{$g$}
\end{picture}
\end{center}
\caption[]{\it Typical Feynman diagrams for associated $\vkk$ production with
$Q',Q=t,b$ quarks at the LHC: the four--body $gg \to Q\bar Q Q' \bar Q'$ 
(a) and the three--body $gb \to b Q\bar Q$ (b) processes.} 
\protect\label{fig:tTbB}
\vspace*{0mm}
\end{figure}
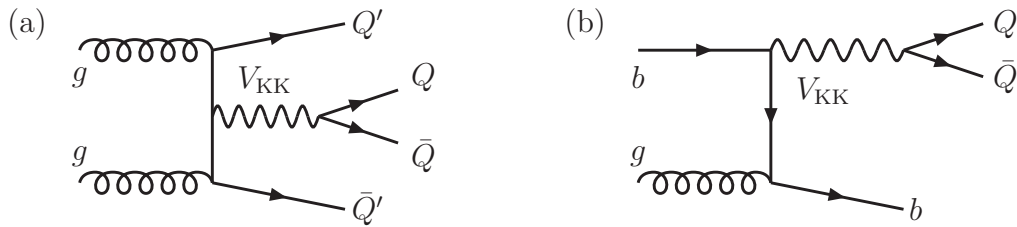

In this section, we briefly discuss the associated production of the KK
excitations of the gauge bosons with  heavy quarks  
\begin{eqnarray}  gg / q
\bar q &\to & Q'\bar{Q'} \ \vkk  \to Q'\bar{Q'} \ Q\bar{Q} \label{eq:f4}\\  
gb &\to& b \ \vkk  \to b \ \ Q\bar{Q} \label{eq:f3}  
\end{eqnarray}  
where $q$ stands for the light quarks present in the proton and $Q/Q'$ stand  
for the final state heavy $t$ and/or $b$ quarks. Some generic Feynman diagrams
are shown in Fig.~\ref{fig:tTbB}. In the first case, a pair of $b$ or $t$ quarks
is produced mainly in the $gg$ fusion subprocess with a small contribution from
$q\bar q$ annihilation, followed by the emission of a KK gauge boson from one of
the heavy quark lines, while in the second case a KK gauge gauge boson is
emitted  from the $b$--quark line in $gb$ fusion with the initial $b$--quark
taken from the proton sea.  In both cases the produced KK gauge  boson decays
predominantly into heavy  quark pairs, leading  to topologies with multi $b$
and/or $t$ final states. \s

In the $2 \to 4$ body reaction of eq.~(\ref{eq:f4}), since the SM production of
quark pairs is dominated by the $gg$ fusion mechanism, the largest contribution
to associated production of KK gauge bosons with heavy quarks comes also from
the same channel. After radiation from the heavy quark line, $\vkk $ can either 
decay into a heavy quark pair of the same flavour as it was produced in
association  with, or of the other heavy flavour. To avoid the combinatorial
background, we  choose to look at $t\bar t + b\bar b$ final states which
originate from the two channels $pp \to b\bar b \vkk $ with $\vkk  \to t\bar t$
and  $pp \to t\bar t \vkk $ with $\vkk  \to b\bar b$.  Since the $ gg \to b\bar
b$  cross section is much larger than the $pp \to t\bar t$ cross section and
that the $t\bar t$ branching  ratio of $\vkk$  is  in general larger than the
$b\bar b$ branching ratio, the final $t\bar t \, b\bar  b$   sample from the
signal is dominantly originating from the subprocess $pp \to b\bar b \vkk  \to
b\bar b \ t\bar t$.\s

As for the kinematics of this production channel, the KK gauge boson is produced
almost  at rest owing to its large mass and hence decays into top quarks with a
wide  range of transverse momenta with an excess of events with large $p_T$.
On the other hand, the $b$ quarks produced in the hard
process of  $gg$ fusion dominantly have small transverse momenta. The picture is
reversed for the second channel $pp \to t\bar t \vkk  \to t\bar t b\bar b$  which,
as already mentioned, gives a  smaller contribution to the signal cross section.
For the QCD background that we have considered, $pp \to t\bar t b\bar b$, and
that we have calculated using the code {\tt CompHEP} \cite{comphep}, the
dominant contribution comes from the subprocesses $gg/q \bar q \to t\bar t$
production followed by the emission of a virtual gluon splitting into a $b\bar
b$ pair. To reduce this QCD background without significantly affecting  the 
signal, we impose  the following set of kinematical cuts
\begin{equation} p_T^{t,\bar t}
\ge 300 \ {\rm GeV}, \ \ p_T^{b,\bar b} \ge 100 \ {\rm GeV}, \ \  |\eta_{Q,\bar
Q}| \le 2 \ . \label{eq:4cut} 
\end{equation}
We have used large cuts on the top quarks transverse momenta to reduce the 
background as much as possible; one can further optimize these cuts to improve 
the significance of the signal, but as the number of events will turn out  to be
rather small,  we do not perform that analysis here. Additionally, we
impose an opening angle cut $\theta_{b\bar b} \ge 60^\circ $ which reduces the
gluon splitting into $b\bar b$ in the background by more than an order of
magnitude and does not affect the signal much.\s

The cross-sections for $t\bar tb\bar b$ production are shown in 
Table \ref{tab:c34} in all four RS scenarios $E_1$ to $E_4$. The numbers for the
RS model correspond to the signal only and do not include the background events
or the interference with the background amplitudes [which should occur in the
case where the KK excitation of the gluon is produced]. The rather low cross
sections for signal events, compared to a SM background of $\sigma^{\rm SM} \sim
0.4$ fb, indicate that even with a high luminosity of ${\cal L}=100$ fb$^{-1}$,
one will not be able to reach a significance of $5\sigma$. The best
significance, $\sim 4 \sigma$, is obtained in the case of scenario $E_2$. Hence,
no discovery is possible with  the $t\bar tb\bar b$ channel of associated
production, if a reasonable luminosity is assumed. We expect the trend for the
other topologies to be similar. Note that here, we considered only the
production of the KK excitation of the gluon; the significance for the
production of the KK excitations of the electroweak gauge bosons [and in
particular of the  $Z^{\prime (1)}$ which cannot be produced in the Drell--Yan
processes discussed in section 3] is even smaller.\s

\begin{table}[!h]
%\vspace*{5mm}
\renewcommand{\arraystretch}{1.3}
\begin{center}
\begin{tabular}{|c|ccc|}\hline
	& $\sigma^{\rm RS}(pp \to t\bar tb\bar b)$ & $\ \sigma^{\rm RS}/\sigma^{\rm SM}$  \ & $\ 
	{\cal S}_{100}^{\rm ttbb}\ $ \\ \hline
$E_1$  &0.06& 0.13 & 0.8 \\
$E_2$  &0.26& 0.58 & 3.9 \\
$E_3$  &0.04& 0.08 & 0.6 \\
$E_4$  &0.16& 0.35 & 2.3 \\  \hline
\end{tabular}
\end{center}
\vspace*{-1mm}
\caption{The signal cross section [fb], the signal to background ratio and the
signal statistical significance for ${\cal L}=100$ fb$^{-1}$ for the four--body 
process $pp \to t\bar tb\bar b$ in the  RS scenarios $E_{1,2,3,4}$ after 
the cuts discussed in the text are applied.}
\label{tab:c34}
\vspace*{-1mm}
\end{table}

Next we turn to the associated production of the KK excitations with a
$b$--quark in  $gb$ fusion. Though the density of $b/\bar b$ quarks in the
proton is  small, the large coupling of the $b$--quark with the KK gluon for
instance could partially compensate for it. Looking at the reaction $gb \to b
g^{(1)} \to b t  \bar t$, where $b$ stands for both the $b$ and the anti-$b$
quark, we use the same kinematical cuts as those introduced in
eq.~(\ref{eq:4cut}). In fact, because now we are dealing with a $2\to 3$
production process, the signal  cross sections are expected to be one order of
magnitude larger than that of the $2\to 4$ process discussed previously. Here
again, the $b$ quark produced in the hard process has low transverse momentum,
while the top quarks  can have large  transverse momenta in the signal events.
The irreducible QCD background $bg \to b\, t\bar t $, that we calculate  again
using {\tt CompHEP} \cite{comphep}, is also more than one order   of magnitude
larger compared to the previous process, $\sigma^{\rm SM} \sim 9$ fb.\s 

The signal production cross sections for a top quark pair and a $b$--type jet
with the kinematical cuts of eq.~(\ref{eq:4cut}) is shown in  of
Table \ref{tab:c35} for the pure RS signal in the  four selected scenarios
$E_{1,..,4}$. Again the points $E_2$ and $E_4$ have large signal  cross sections
owing to the large coupling of the $b$ quarks to the KK gluon. For the other two
points, the couplings of $b$ quark are small and, hence, lower production rates
of the KK gluon are obtained. With an integrated luminosity of ${\cal L}=100$
fb$^{-1}$, one obtains a significance which is larger than 8 for scenarios $E_2$
and $E_4$, while it is about 3 or less for the other two scenarios. We again
have considered only the associated production of a KK gluon and not accounted
for the interference between the RS signal and the background.

\begin{table}[!h]
\vspace*{2mm}
\renewcommand{\arraystretch}{1.3}
\begin{center}
\begin{tabular}{|c|ccc|}\hline
	& $\sigma^{\rm RS}(pp \to t\bar t b/t\bar t\bar b)$ & $\sigma^{\rm RS}/\sigma^{\rm SM}$  & ${\cal 
	S}_{100}^{\rm ttb}$ \\ \hline
$E_1$  &0.91 & 0.10 & 3.0  \\
$E_2$  &4.12 & 0.47 & 13.8 \\
$E_3$  &0.61 & 0.07 & 2.0 \\
$E_4$  &2.50 & 0.28 & 8.4 \\  \hline
\end{tabular}
\end{center}
\vspace*{-1mm}
\caption{The signal cross section [fb], the signal to background ratio and the signal
statistical significance for ${\cal L}=100$ fb$^{-1}$ for the three--body 
process $gb \to t\bar t b/t\bar t\bar b$ in the four RS scenarios $E_{1,2,3,4}$ after 
the cuts discussed in the text are applied.}
\label{tab:c35}
\vspace*{-1mm}
\end{table}

Thus, at least in the 3--body production process $g b \to b t\bar t$, the cross
section in some RS scenarios can  be significantly larger than the QCD 
background cross section. However, to reach a signal significance which is
important, large integrated luminosities, ${\cal L} \gsim 100$ fb$^{-1}$, are 
needed. In this case, 
the top and bottom quark jets cannot
be identified with an excellent efficiency and more background processes with
potentially much larger rates need to be considered. Detailed experimental
analyzes need thus to be performed in order to assess the viability of this
signal. 

\subsection*{5. Conclusions}

In the framework of the Randall--Sundrum extra--dimensional model with SM fields
in the bulk, we have studied the production of the KK excitations of gauge
bosons at LHC which then dominantly decay into heavy top and bottom quark pairs.
For illustration,  we have considered a specific version of the model in which
the SM gauge group is enhanced to a left--right symmetry and concentrated on
quark geometrical localizations  and gauge quantum numbers that allow to solve
the LEP anomaly on the  forward--backward asymmetry $A_{FB}^b$. We have selected
some characteristic points of the parameter space of this model and assumed 
masses $\mkk \sim 3$ TeV for the KK excitations of the gluon and the
electroweak gauge bosons as dictated by constraints from high--precision
electroweak data. \s

We have shown that the contribution of the Drell--Yan process $q\bar q \to  \vkk
\to t\bar t$  to the $pp \to q\bar q/gg \to  t\bar t$ cross section, in which 
we have taken into account the total width of the excitation as well as the 
interference between the signal and SM the background, can be substantial and
would allow for the detection of the gluon KK excitation with a high statistical
significance; the signal significance  is smaller for the excitations of the
electroweak gauge bosons as a result of the smaller couplings.  Besides the
invariant mass distributions and the sizeable total widths of the KK resonances,
the polarization of the top quark and the forward--backward asymmetry, would
allow to characterize the signal and to probe the chiral  structure of the
fermion couplings of the KK excitations. In the case of the $q \bar q \to  \vkk
\to b \bar b$ process, a large signal significance compared to the SM background
is also obtained.\s

We have also calculated the cross sections for the higher order processes in
which the KK  excitations couple only to the heavy top and bottom quarks and
compared  them to their  corresponding irreducible SM backgrounds. We have first
studied  the gluon--gluon fusion mechanism, $gg \to \vkk \to t \bar t, b\bar b$,
which is induced  by a heavy quark loop and in which the anomalous $gg \vkk $
vertex has to be  regulated via the St\"uckelberg mixing term. We have then
analyzed  the four--body reactions $pp \to t\bar t b\bar b$,  $t\bar t t\bar t$
and $b\bar b b\bar b$ as well as the three--body reactions $gb \to b t \bar t$
and  $b b\bar b$, in which the KK excitations are radiated from the heavy quark
lines. The cross sections for these higher order processes at the LHC, except
potentially for the $bg \to g^{(1)}b$ process, turn out to be too small 
compared to the SM background, to allow for a detection of heavy  KK excitations
with $\mkk \gsim 3$ TeV. \s

In our parton--level analysis, we took into account only the irreducible  SM QCD
background and did not attempt to include experimental effects such as
detection  efficiencies, $b$--quark tagging, reducible light jet backgrounds
etc.. Only refined and realistic Monte--Carlo analyses, which take into account
all these effects,  would allow to assess the viability of the KK gauge boson
signals. These experimental analyses are currently being performed by some
members of the ATLAS and CMS collaborations \cite{analysisLHC}

\subsubsection*{Acknowledgments}

We thank Fawzi Boudjema  and Emilian Dudas for useful discussions on the anomaly
cancellation in the $gg \to \vkk $ process, as well as Rohini Godbole and
Fran\c{c}ois Richard for useful suggestions. This work is supported by the
CEFIPRA project no.\,3004-B and by the French ANR project PHYS@COL\&COS.

\end{document}